%% file: User_Guide_arxiv.tex
\documentclass[twocolumn, 3p, number, sort&compress]{elsarticle}
\usepackage{amsmath,amssymb,graphicx,hyperref}
\usepackage[utf8]{inputenc}
\usepackage[T1]{fontenc}
\usepackage{multirow}
\usepackage{fancyvrb}
\usepackage{color}
\usepackage[usenames,dvipsnames]{xcolor}
\usepackage{pdflscape,array,booktabs}

\usepackage{siunitx}
\usepackage{matlab-prettifier}
\usepackage{listings}
\usepackage{bm}
\newcommand{\dd}{\ensuremath{\text{d}}}
\newcommand{\MATLAB}{\textsc{Matlab}}
\newcommand{\tmatrix}{\textit{T}-matrix}
\newcommand{\FORTRAN}{\textsc{Fortran}}
\newcommand{\CODENAME}{\textsc{smarties}}
\newcommand{\PACKAGENAME}{\texttt{smarties}}

\title{\CODENAME: user-friendly codes for fast and accurate calculations of light scattering by spheroids}
\begin{document}
\author{W. R. C.~Somerville}
\ead{wrcsomerville@gmail.com}
\author{B.~Auguié}
\ead{baptiste.auguie@gmail.com}
\author{E. C.~Le~Ru\corref{cor1}}
\ead{eric.leru@vuw.ac.nz}
\address{The MacDiarmid Institute for
  Advanced Materials and Nanotechnology, School of Chemical and
  Physical Sciences, Victoria University of Wellington, PO Box 600,
  Wellington 6140, New Zealand}
\cortext[cor1]{Corresponding author}

\onecolumn
\begin{abstract}
We provide a detailed user guide for \CODENAME, a suite of \MATLAB\ codes for the calculation of the optical properties
of oblate and prolate spheroidal particles, with comparable capabilities and ease-of-use as Mie theory for spheres.
\CODENAME\ is a \MATLAB\ implementation of an improved \tmatrix\ algorithm for the theoretical modelling of electromagnetic scattering by particles of spheroidal shape. The theory behind the improvements in numerical accuracy and convergence is briefly summarised, with reference to the original publications. Instructions of use, and a detailed description of the code structure, its range of applicability, as well as guidelines for further developments by advanced users are discussed in separate sections of this user guide. The code may be useful to researchers seeking a fast, accurate and reliable tool to simulate the near-field and far-field optical properties of elongated particles, but will also appeal to other developers of light-scattering software seeking a reliable benchmark for non-spherical particles with a challenging aspect ratio and/or refractive index contrast.
\end{abstract}
\maketitle

\tableofcontents

\twocolumn
\section{Introduction}

We present a user guide and description of \CODENAME, a numerically stable and highly accurate implementation of the \tmatrix{}/Extended Boundary-Condition Method (EBCM) for light-scattering by spheroids, based on our recent work \cite{JQSRT2012,JQSRT2013,JQSRT2015}.
The complete package can be downloaded freely from {http://www.victoria.ac.nz/scps/research/research-groups/raman-lab/numerical-tools}, see Sec.~\ref{Licensing}
for licensing information. The name of the program stands for \emph{Spheroids Modelled Accurately with a Robust \tmatrix\ Implementation for Electromagnetic Scattering}, and is also a nod to the well-known colourful candy of oblate shape.

\subsection{Description and overview}

This package contains a suite of \MATLAB\ codes to simulate the light scattering properties of spheroidal particles, following the general \tmatrix\ framework \cite{2002Mishchenko}. The scatterer should be homogeneous, and described by a local, isotropic and linear dielectric response (this includes metals, but not perfect conductors). Magnetic, non-linear, and optically active materials are not considered. The surrounding medium is described by a lossless, homogeneous and isotropic dielectric medium extending to infinity.

\CODENAME\ specifically implements recently-developed algorithms for numerically accurate and stable calculations. The general EBCM/\tmatrix\ method is described in detail in Ref.~\cite{2002Mishchenko}, while the underlying theory and relevant formulas for our specific improvements are described in Ref.~\cite{JQSRT2013}, with additional information found in \cite{JQSRT2012,JQSRT2015}. The relevant equations and sections from both Ref.~\cite{JQSRT2013} and Ref.~\cite{2002Mishchenko} are referenced when possible as ``inline comments'' to the code.

The package includes detailed examples and can also be used by a non-specialist with an application-oriented perspective, requiring no specific knowledge of the underlying theory.

The package contains:
\begin{itemize}
\item
Six ready-to-run example scripts to calculate standard optical properties, namely: fixed-orientation and orientation-averaged far-field cross-sections, near fields, \tmatrix\ elements, and scattering matrix elements. Examples also cover the simulation of wavelength-dependent spectra of surface-field and far-field properties.
\item
Two tutorial scripts where such simulations are further detailed with step-by-step instructions, exposing the lower-level calculations of intermediate quantities.
\item
Additional \emph{high-level} and post-processing functions, which can be used by users to write new scripts tailored to their specific needs.
\item
A number of \emph{low-level} functions, which are used by the code and might be adapted by advanced users.
\item
Dielectric functions for a few materials such as gold and silver, implemented via analytic expressions \cite{2006EtchegoinJCP,2009Book} or silicon, interpolated from tabulated values.
\end{itemize}

\subsection{Relation to other codes}

Standard \tmatrix/EBCM codes in \FORTRAN\ have already been developed \cite{1990Barber,2005QuirantesJQSRT}, with those by Mishchenko and co-workers \cite{1998MishchenkoJQSRT} arguably the most popular. These freely-available codes provide a wide range of capabilities (including for example different particle shapes) and have been widely used and tested. The standard EBCM method however suffers from a number of numerical problems and instabilities for large multipole orders, which are necessary for either high precision, large particles, elongated particles, near-field calculations, or any combination of the above. This can result in inaccurate results and in some cases in complete loss of convergence. This unreliable behaviour for numerically challenging simulations can make the method difficult to use for non-experts, who may find it hard to ``tune'' the parameters that ensure accuracy and convergence. It also impedes the theoretical study of the intrinsic convergence properties of the \tmatrix\ method, obfuscated by (implementation-dependent) numerical loss of precision \cite{JQSRT2015}.

Recently, we have identified the primary causes for numerical instabilities in the special (but important) case of spheroidal particles \cite{JQSRT2012} and proposed a new algorithm to overcome them \cite{JQSRT2013}. Thanks to those improvements, high accuracy and reliable convergence can be obtained over a wider range of parameters, especially toward high aspect-ratio (elongated) particles where the standard EBCM implementation would fail \cite{JQSRT2015}. This document aims to present and discuss a publicly available \MATLAB\ implementation of these recent developments. Our package should complement, rather than replace, existing \tmatrix\ codes such as those of Mishchenko \cite{1998MishchenkoJQSRT}. The present code offers a number of advantages:
\begin{itemize}
\item
Thanks to the improvements in accuracy and convergence, we believe this code will be readily accessible to non-expert users and allow the routine calculation of optical properties of spheroids as easily as with Mie theory for spheres. An example is provided in Section~\ref{sec:studies} as a demonstration.
\item
We also provide specific routines to compute near fields and surface fields, which will be beneficial to the exploitation of this powerful method in areas such as nanophotonics, optical trapping, plasmonics, etc., where the \tmatrix/EBCM method has not been widely applied.
\item
\MATLAB\ provides an easy-access, interactive environment to carry out a broad range of numerical simulations, and plot/export the results conveniently.
\item
The accuracy of the obtained results can be easily estimated for any type of calculation, owing to the well-behaved convergence of the improved algorithm.
\item
A wider range of parameters can be simulated, especially scatterers with large aspect ratios.
\end{itemize}
A number of limitations should be also be noted:
\begin{itemize}
\item
These codes are limited to spheroidal particles, for which we identified and circumvented numerical problems that are very specific to this geometrical shape.
\item
\MATLAB\ is inherently slow compared to compiled languages such as C or \FORTRAN, which may be an issue for intensive calculations (for example the simulation of polydisperse samples, with particles varying in size and shape). We envisage that this implementation could serve as a template for a future port of this new algorithm to a more efficient language.
\item
The calculation of some derived properties, e.g. the scattering matrix, has not been optimized and could be particularly slow.
\item
Although the range of parameters that may be simulated with reasonable accuracy has been extended toward larger
aspect ratios, the method is still limited to moderate particle sizes; and even small sizes only for particles with
a large relative refractive index. In this case, the matrix inversion step is the limiting factor and extended-precision
arithmetic as implemented in \cite{1998MishchenkoJQSRT} would be required to overcome it.
\end{itemize}

\subsection{Aims of this manual}

This document was written with two types of users in mind:

\begin{itemize}
  \item
Researchers interested in simulating electromagnetic scattering by nonspherical particles for practical applications, and seeking an efficient and (relatively) fool-proof program with ease of use comparable to Mie theory.
  \item
   Other developers of electromagnetic scattering software interested in benchmarking calculations against a highly-accurate reference.
\end{itemize}
With this dual perspective, we have divided the source code into \emph{low-level} and \emph{high-level} functions, including complete scripts for specific calculations, but also documented how to access intermediate quantities such as the \tmatrix\ elements. This user guide is also divided into sections that reflect these two complementary objectives, with Sections \ref{sec:principles} and \ref{sec:implementation} focusing on more theoretical aspects and in-depth description of the code implementation.

\subsection{Licensing}
\label{Licensing}

\CODENAME\ is licensed under the Creative Commons Attribution-NonCommercial 4.0 International License. To view a copy of this license, visit \url{http://creativecommons.org/licenses/by-nc/4.0/}.

The package, including all its files and content are under the following copyright: 2015 Walter Somerville, Baptiste Auguié, and Eric Le Ru. The package may be used freely for research, teaching, or personal use. The unmodified complete package may be re-distributed and freely exchanged for academic research or government use, but cannot be commercialized or used for commercial purposes. The theory and code should be appropriately referenced by citing this user-guide in any presentation of results obtained using this package (or any other code using it).

\subsection{Disclaimer}

These codes have been developed and tested with \MATLAB\ 7.14 (R2012a) \cite{MATLAB:2012}, GNU Octave 4.0.0 \cite{octave} (open-source software) and \MATLAB\ 8.5 (R2015a) on a PC running Microsoft Windows 7 x64. The code is also known to run under MacOS X (10.10) and Linux (Ubuntu 15.04). Slight changes may be necessary to run them on older (or newer!) versions of \MATLAB/Octave.

Although every effort has been made to get rid of bugs (programming bugs, or incorrect physical formulas) and to test the code against existing ones, some issues may still be present. We hope the users will help us identify them and we will try to update the code when necessary.

Note also in this context that these codes do not implement a strict check of user input; if incorrect parameters are passed in a function call, errors will occur.

The authors do not accept any responsibility for improper use of the program, accidental errors that may still be present, or improper interpretation of its limitations and/or results derived therefrom. It is the responsibility of the user to check the validity of the inputs/outputs, their physical interpretation, and their suitability for her/his specific problem.

\subsection{Feedback}

We would like to hear from the users of this code to improve it over time. This feedback could include simple issues of layout and organization of the information or plain errors. Please feel free to send us any feedback (good or bad), bug reports, questions, comments, or suggestions to \url{eric.leru@vuw.ac.nz}.

%

\section{Getting started}

\begin{figure*}[!htpb]
  \includegraphics[width=\textwidth]{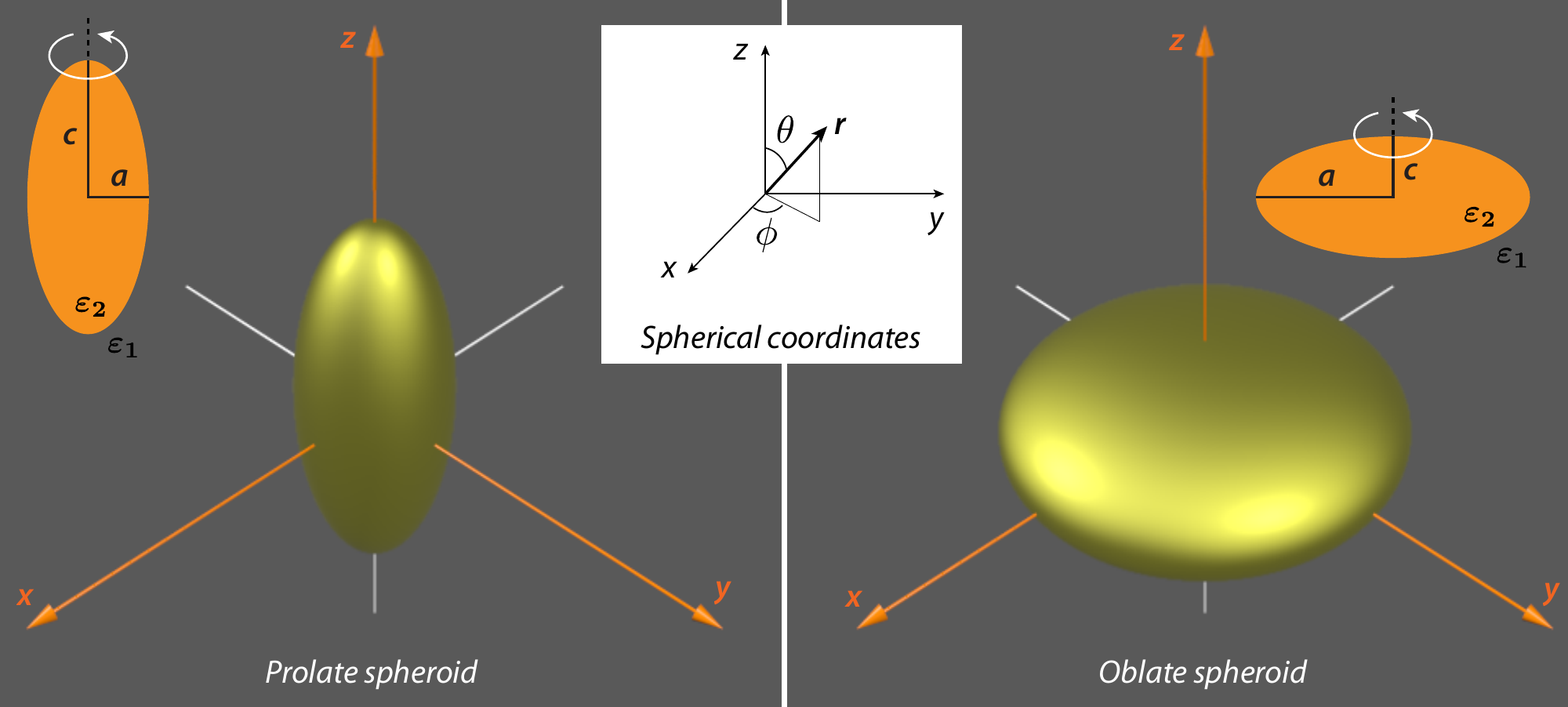}
  \caption{3D illustation of spherical coordinates, and the geometrical parameters for prolate (left) and oblate (right) spheroids. The axis of revolution is along $z$.}
  \label{fig:schematic}
\end{figure*}

Figure~\ref{fig:schematic} depicts the scatterer's geometry: the spheroid, either prolate or oblate, has a fixed orientation with symmetry axis along $z$. For completeness, analytical formulas related to the geometry of spheroids are provided in section~\ref{sec:geometry}. The incident field may be specified along an arbitrary direction, defined in spherical coordinates. The full set of parameters required to run a simulation, including the specification of the incident field, is defined in section~\ref{sec:parameters}.

\subsection{Installation}

\begin{itemize}
\item
Download the \CODENAME~package from our webpage \cite{codeurl}.
\item
Unzip the \PACKAGENAME\texttt{.zip} file, keeping the subdirectory structure for clarity.
\item
Set your \MATLAB\ current directory to \PACKAGENAME\ and
  run the \texttt{InitPath.m} function once in \MATLAB\ to add all the subdirectories to your \MATLAB\ search path. All the functions and scripts are then accessible from the \MATLAB\ command line. This allows all codes to run and communicate with each other irrespective of the current directory.
\end{itemize}

Note that you must run \texttt{InitPath.m} each time you restart \MATLAB. To avoid this step, you may add all \PACKAGENAME\ folders to your \MATLAB\ path permanently or edit the \texttt{startup.m} file to do that (check \MATLAB's help for details).

\subsection{Octave users}

Octave is an open-source alternative to \MATLAB. Although our codes were mostly developed and tested for \MATLAB, we have tried to ensure their compatibility with Octave. Most scripts should therefore run as-is with Octave, although they were typically slower in our tests. The numerical accuracy may also differ, as well as the rendering of graphics.

\subsection{Initial steps}

The easiest way to get started is to run one of the example files in the \texttt{Scripts} folder (those starting with \texttt{ScriptSolve}) and change the parameters as needed. These scripts provide a direct example of how to call a number of high-level functions designed to solve specific problems of interest. These are
\begin{itemize}
\item
\texttt{ScriptSolveForT}: Calculates the \tmatrix, scattering matrix, and orientation-averaged properties for a single wavelength.
\item
\texttt{ScriptSolveForFixed}: Calculates the field expansion coefficients and the corresponding far-field cross-sections for a fixed-orientation, for a single wavelength.
\item
\texttt{ScriptSolveForSurfaceField}: Calculates the field expansion coefficients and the corresponding far-field cross-sections and surface fields for a fixed-orientation.
\item
\texttt{ScriptSolveForTSpectrum}: Calculates the \tmatrix\ and the orientation-averaged properties for multiple wavelengths.
\item
\texttt{ScriptSolveForFixedSpectrum}: Calculates the field expansion coefficients  and the corresponding far-field cross-sections for a fixed-orientation, as a function of wavelength.
\item
\texttt{ScriptSolveForSurfaceFieldSpectrum}: Calculates the field expansion coefficients and the corresponding far-field cross-sections and surface fields  for a fixed-orientation, as a function of wavelength.
\end{itemize}

Those scripts define the parameters of the simulation, call the corresponding high-level functions to perform the calculations, and output the most important results in the \MATLAB\ console and/or as interactive graphics. Convergence tests are also performed as part of the calculations, and accuracy estimates for the results are included in the displays.

In order to understand in more detail how the code operates, we also provide two example scripts, \texttt{ScriptTutorial} and \texttt{ScriptTutorialSpectrum},  where all the main steps in the calculation are listed explicitly with extensive comments about the meaning of the various parameters. We recommend copying and editing these example scripts to solve user-specific problems and/or implement custom extensions to the current code.

Most functions start with a detailed help and are commented within the code. Typing \texttt{help FunctionName} will display the corresponding help information.

The information below summarizes and complements the inline comments included in the six example scripts \texttt{ScriptSolve...}. It provides the most important technical details of the implementation, for users wishing to write additional custom routines.

\subsection{Definition of the parameters}
\label{sec:parameters}

For the calculation of the \tmatrix\ and orientation-averaged cross-sections, only four parameters are needed to define the scatterer properties:
\begin{itemize}
\item
\texttt{a}: semi-axis along $x,y$.
\item
\texttt{c}: semi-axis along $z$ (axis of rotational symmetry).
\item
\texttt{k1}: wavevector in embedding medium (possibly a wavelength-dependent vector).
\item
\texttt{s}: relative refractive index $s$ (possibly a wavelength-dependent vector).
\end{itemize}

Note that consistent units must be used for $a$, $c$, and $k_1$, e.g. $a$ and $c$ in nm and $k_1$ in nm$^{-1}$.

$k_1$ denotes the wavevector outside the particle, where the refractive index is $n_1=\sqrt{\varepsilon_1}$ (assumed real positive):
\begin{equation}
k_1=\frac{\omega}{c}\sqrt{\varepsilon_1}=\frac{2\pi}{\lambda}\sqrt{\varepsilon_1}.
\end{equation}
The relative refractive index (adimensional) is defined as
\begin{equation}
s=\frac{\sqrt{\varepsilon_2}}{\sqrt{\varepsilon_1}}.
\end{equation}
Both $s$ and $\varepsilon_2$ may be complex (for absorbing and/or conducting particles).\\

The $P,Q,T,R$-matrices computation requires the following parameters:
\begin{itemize}
\item
\texttt{N}: Number of multipoles $N$ requested for the \tmatrix\ (and $R$-matrix).
\item
\texttt{nNbTheta}:  Number of angles $\theta$ used in Gaussian quadratures for the evaluation of $P$- and $Q$-matrix integrals.
\end{itemize}
The function \texttt{sphEstimateNandNT} may be used to automatically estimate those latter two parameters for best convergence,
but we we nevertheless recommend that the convergence of the calculations be checked to ensure reliable results.\\

For convenience those six parameters may be collated in a \emph{structure} (a \MATLAB\ object akin to a list, called \texttt{stParams} in our example scripts), which is passed to the high-level (\texttt{slv...}) functions.

Additionally, one of the following two parameters is needed if the field expansion coefficients and/or the cross-sections for a given fixed orientation are sought:
\begin{itemize}
\item
\texttt{sIncType}: String defining the type of incident plane wave, e.g. \texttt{\textquotesingle KxEz\textquotesingle} for a wave incident along $x$ and linearly polarized along $z$. This shorthand notation is only defined for a few standard combinations, namely \texttt{KxEz, KxEy, KyEz, KyEx, KzEx, KzEy}. In other cases, use \texttt{stIncPar}.
\item
\texttt{stIncPar}: Structure defining a linearly-polarized incident plane wave excitation via three Euler angles. It can be obtained from calling \texttt{vshMakeIncidentParameters}.
\end{itemize}

For field calculations (such as surface fields), further parameters are required:
\begin{itemize}
\item
\texttt{nNbThetaPst}: Number of angles $\theta$ for post-processing (should typically be larger than \texttt{nNbTheta} for accurate surface averaging).
\item
\texttt{lambda}: Wavelength (in free space) [in the same unit as $a$, $c$, $k_1^{-1}$].
\item
\texttt{epsilon2}: Dielectric function $\varepsilon_2$ of scatterer (possibly complex).
\item
\texttt{epsilon1}: Relative dielectric constant $\varepsilon_1$ of embedding medium (real positive).
\end{itemize}
Note that the latter three are not independent of $k_1$ and $s$ that have already been defined. Those additional parameters should also be included in \texttt{stParams}.

Finally a number of optional settings can also be defined in a structure \texttt{stOptions}:
\begin{itemize}
\item
\texttt{bGetR}: Boolean (default: \texttt{false}). If false, the $R$-matrix and internal field coefficients are not calculated. The default value will be overridden by functions requiring $\mathbf{R}$.
\item
\texttt{Delta}: Number of extra multipoles for $P$- and $Q$- matrices, i.e. $N_Q = N + \Delta$. Default is $\Delta=0$.
If \texttt{Delta=-1}, then the code tries to estimate it from the convergence of $T^{22,m=1}_{11}$ (see \cite{JQSRT2015} for details),
by calling \texttt{sphEstimateDelta}.
\item
\texttt{NB}: Number of multipoles to compute the Bessel functions in the improved algorithm ($N_B \geq N_Q$). If \texttt{NB=0}, then $N_B$ is estimated by calling \texttt{sphEstimateNB}, which is the case by default.
\item
\texttt{absmvec}: Vector containing the values of $|m|$ for which $\mathbf{T}$ is to be computed. These values are limited to $0\leq |m| \leq N$. To compute all $m$ (most cases of interest), simply use \texttt{absmvec=0:N} (which is the default value).
\item
\texttt{bGetSymmetricT}: Boolean (default: \texttt{false}). If true, $\mathbf{T}$ is symmetrized as described in Sec.~\ref{Tcomp}.
\item
\texttt{bOutput}: Boolean (default: \texttt{true}). If false, suppresses some of the output printed in the \MATLAB\ console, which is a better option for example in calculations of spectra with many wavelengths.
\end{itemize}

\subsection{Minimal example}
The following script is set up to simulate the far-field cross-sections of a gold prolate spheroid in air, at a single wavelength $\lambda=650$\,nm. The simulation parameters are stored in a structure \texttt{stParams} for convenience. Only one optional parameter is defined in \texttt{stOptions} (for the others, default values will be used). These two structures are passed to the high-level \texttt{slvForFixed} function that implements the calculation of the expansion coefficients and cross-sections for a fixed orientation.

\begin{lstlisting}[
  frame=tb,
  style      = Matlab-Pyglike,
  basicstyle = \footnotesize\mlttfamily,
  title={\emph{Minimalist script showing how to set up a simulation}}
  ]
% Parameters of the scattering problem
% stored in a structure
lambda = 650;
stParams.a=10; stParams.c=50;
stParams.k1=2*pi/lambda;
stParams.s=sqrt(epsAu(lambda));
stParams.sIncType = 'KxEz';

% Optional control parameters
stOptions.bGetR = false;

% Automatically estimates required N and nNbTheta
[stParams.N, stParams.nNbTheta] =
   sphEstimateNandNT(stParams,stOptions);

%% T-matrix calculation and output of cross-sections
stC = slvForFixed(stParams, stOptions);
sprintf('Cext = %g, Cabs = %g, Csca = %g',
            stC.Cext, stC.Cabs, stC.Csca)
\end{lstlisting}

This general structure is followed by all the examples, with varying levels of complexity, and the high-level functions such as \texttt{slvForFixed} performing the actual calculations are grouped in the \texttt{Solve} directory.

\subsection[Convergence, range of validity]{Convergence, accuracy, and range of validity}
\label{sec:convergence}

One of the problems of the conventional \tmatrix/EBCM method is to study its convergence and accuracy. This is because the method becomes unstable with multipoles of high order, which may occur before the results have fully converged. Many of those issues have been solved in the present method, as discussed in Ref.~\cite{JQSRT2015}. Ref.~\cite{JQSRT2015} also provides a detailed discussion of the parameters affecting convergence and accuracy.

Thanks to the improved stability, we propose a simple and reliable convergence and accuracy test that will work in most cases. It consists in repeating the same calculations with a larger number of multipoles $N$ and quadrature points $N_\theta$, for example: $N'=N+5$ and $N'_\theta = N_\theta+5$. If surface-averaged properties are calculated, the number of quadrature points used in post-processing should also be checked independently.

In our experience, this simple convergence test provides a reliable estimate of the accuracy of the results. It is implemented in the six example scripts provided with the code.

Obviously, such a test will double the required computing times; for repeated computations such as spectra with many wavelengths, we therefore recommend to only test the most numerically-challenging cases, typically the largest size parameter and/or largest value of $|s|$.

The function \texttt{sphEstimateNandNT} can be use to estimate automatically the required $N$, $N_\theta$ for a simulation. This function should not replace the convergence test described above as it only relies on the convergence of the orientation-averaged extinction cross-section (and only for $m=0,1$) and may fail in rare cases. It does however provide a good first guess for those parameters, and can in addition be used to study how they depend on the scatterer properties, or test the range of validity of the method.

An example of such results is given in Tables~\ref{tab:oblate} and \ref{tab:prolate} of the Appendix for oblate and prolate spheroids, respectively, where the required $N$ and $N_\theta$, along with the obtained accuracy, are summarized as a function of maximum size parameter and aspect ratio for $s=1.311$. Interestingly, when expressed in terms of the maximum size parameter, $x_\mathrm{max}=k_1\mathrm{max}(a,c)$, almost identical convergence requirements were obtained for oblate and prolate spheroids. From those tables, we also infer that the accuracy and stability do not depend strongly on aspect ratio (in stark contrast with the standard EBCM, which rapidly becomes unstable for larger aspect ratios). There remains however an upper limit on the size of particles that can be modeled, which is comparable to the upper limit of double-precision implementations of the standard EBCM at low aspect ratio \cite{1998MishchenkoJQSRT}.

\newcommand\labelxmaxobl{\label{tab:xmaxobl}}
\newcommand\labelxmaxpro{\label{tab:xmaxpro}}
\input{summaryXmaxObl.tex}

Additional automatic tests were carried out to estimate the maximum computable size parameter for a given $h$ and $s$. Those results are summarized in Table~\ref{tab:xmaxobl} for oblate spheroids, with the corresponding table for prolate spheroids in Appendix Table~\ref{tab:xmaxpro}. These computer-generated estimates provide an overview of the range of validity of this new implementation. These suggest that, as a rule of thumb, the method will start to fail when the maximum size parameter $x_\mathrm{max}=k_1 \text{max}(a,c)$ approaches the limits $|s| x_\mathrm{max} \approx 50$ for relatively low aspect ratios, progressively going down to $|s| x_\mathrm{max} \approx 30$ for the largest aspect ratios.
For relatively large aspect ratios, for example $h=20$, the upper limit of size parameter therefore becomes comparable to \emph{extended-precision} implementations of the standard EBCM ($x_\mathrm{max}\approx 32$ for oblate spheroids with $s=1.311$ \cite{1998MishchenkoJQSRT}). As for the standard EBCM codes, a large relative index $|s|$ however remains very challenging.

Also notable from these tables is the fact that a very large number of Gaussian quadrature points are necessary for large aspect ratios of any size. This can be explained from the high curvature of the tip around $\theta=0$ and suggests that much more efficient quadrature schemes could be developed for those cases, e.g. simply using subdivisions of the range of integration with different density of points.

\subsection[Case study: Ag spheroids]{Case study: influence of size and aspect ratio on the far-field properties of silver spheroids}
\label{sec:studies}

\begin{figure}[!htpb]
  \includegraphics[width=\columnwidth]{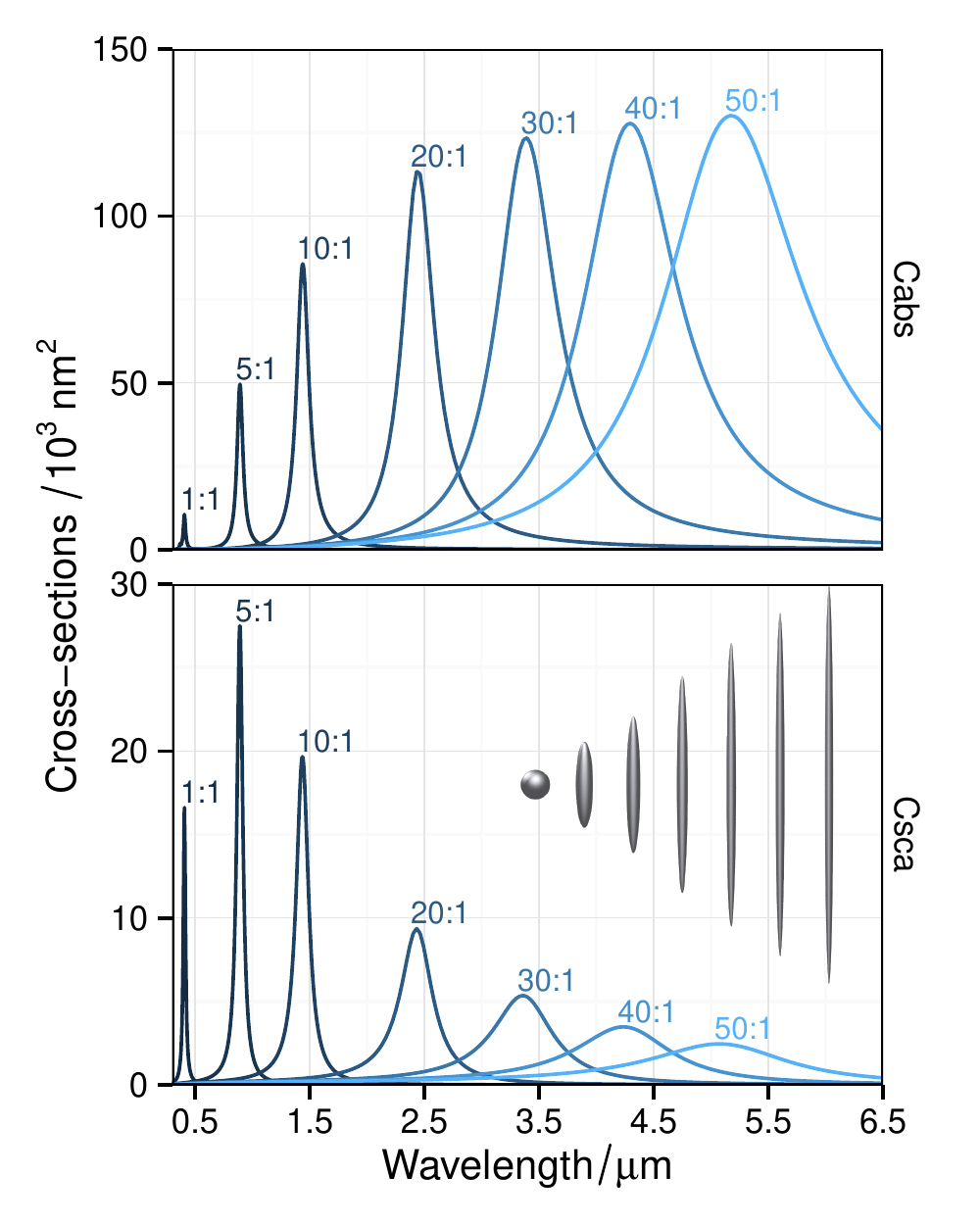}
  \caption{Example calculation of scattering and absorption spectra of prolate Ag spheroids in water with varying aspect ratio $h$ (1 to 50), with a fixed equivalent-volume radius $r_V=20$\,nm.}
  \label{fig2}
\end{figure}

We now illustrate the ease of use of this code for realistic, application-oriented calculations, with a comprehensive example simulating optical spectra of prolate silver spheroids, as a function of size and aspect ratio. The calculation is fully automated with built-in precision checking to ensure a minimum relative accuracy of $10^{-3}$ (better precision could easily be obtained, at the cost of increased computation time).
Figure~\ref{fig2} illustrates this calculation for prolate spheroids with challenging aspect ratios, up to 50:1.

\begin{figure}[!htpb]
  \includegraphics[width=\columnwidth]{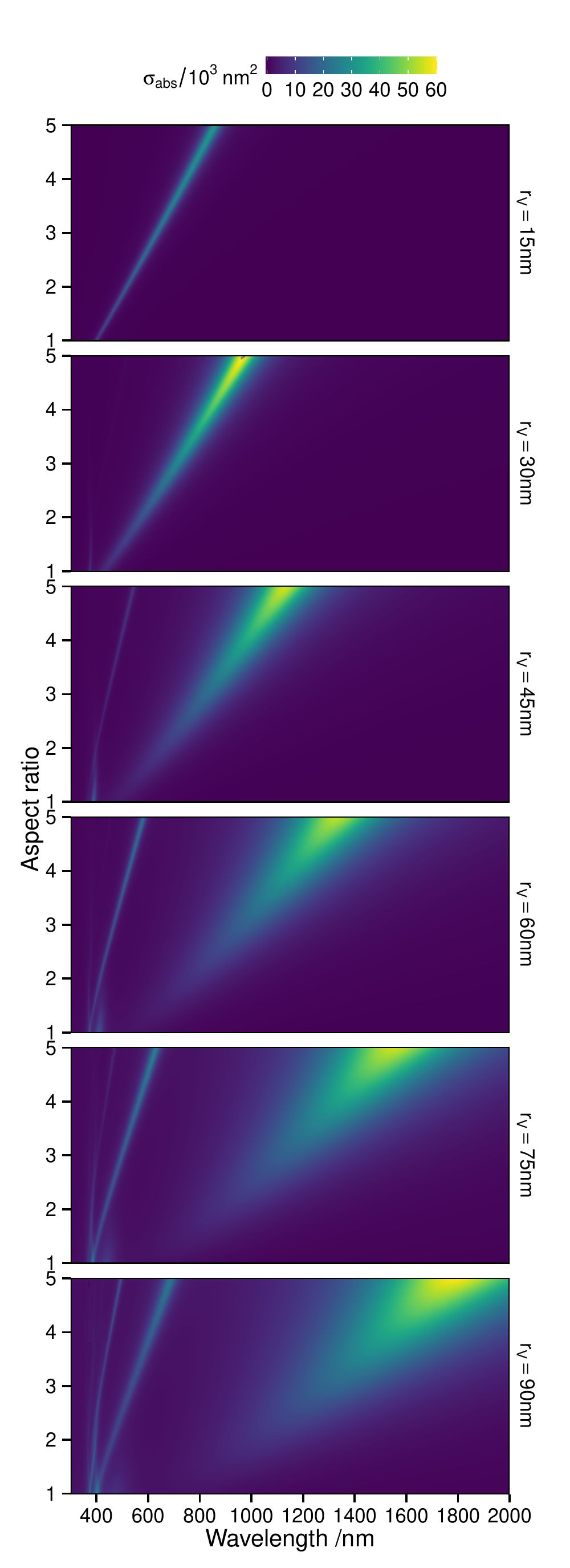}
  \caption{Colour map of fixed-orientation absorption spectra of Ag spheroids in water with varying aspect ratio (from 1 to 5), for 6 different sizes (equivalent-volume radius $r_V$ varied from 15\,nm to 90\,nm, from top to bottom).}
  \label{fig3}
\end{figure}

Another set of results is presented in figure \ref{fig3}, which provides a comprehensive perspective on the optical properties of such particles, with sizes closer to experimentally-accessible values. The optical response is dominated by plasmon resonances, which vary with the size and shape of the particles.
The full calculation presented in figure \ref{fig3} ran for a few hours on a standard desktop computer, and produced $\sim  3.6\times 10^5$ data points (1200 combinations of parameter values, and 300 wavelengths per spectrum). A separate example is also included in that directory for surface-field calculations.

Figure~\ref{fig3} highlights a number of interesting physical features of relevance to the field of plasmonics. Small elongated particles behave as nano-antennas, with a dipolar resonance that red-shifts with increasing aspect ratio \cite{2009BoyackPCCP}. The strength of the absorption increases initially with larger particle size, and red-shifts, but as the larger particles scatter more efficiently the plasmon resonance suffers additional radiative damping, which results in a broadening of the resonance, and a plateau of peak absorption at larger sizes. Larger particle sizes also support multipolar resonances, here visible in the bottom two panels as sharp lines around 400--600\,nm.

Surface fields at specific points, or averaged over the whole particle surface, may also be calculated with similar ease. The ability to routinely simulate the electromagnetic response of elongated particles with comparable ease of use and accuracy to Mie theory should thus enable, as demonstrated in this illustrative example, the exploration of a much broader range of parameters, and perhaps bring out new physical insights.

\section{Underlying principles of the code}
\label{sec:principles}

A detailed description of the \tmatrix/EBCM method can be found in Ref.~\cite{2002Mishchenko}. The most important aspects and notations along with the details of the new algorithm implemented here can be found in Ref.~\cite{JQSRT2013}. We will not repeat all this information here, only summarize the most relevant
aspects, and will refer to the equations and notations of Ref.~\cite{JQSRT2013} when needed.

\subsection{Spherical coordinates}

To apply the \tmatrix/EBCM method, the geometry must be defined in spherical coordinates, with the following conventions (see inset of Fig.~\ref{fig:schematic}): a point $M$ is represented by $(r,\theta,\phi)$ where,
\begin{itemize}
\item
$r \ge 0$ is the distance from origin O.
\item
$0\le \theta \le \pi$ is the co-latitude, angle between $\mathbf{e_z}$ and $\mathbf{OM}$.
\item
$0 \le \phi \le 2\pi$ is the longitude, angle between $\mathbf{e_x}$ and the projection of $\mathbf{OM}$ on ($x$O$y$).
\end{itemize}

The spherical coordinates are thus related to the Cartesian coordinates by:
\begin{equation}
\left\{
\begin{aligned}
x &= r \sin\theta \cos \phi \\
y &= r \sin \theta \sin \phi \\
z &= r \cos \theta
\end{aligned}
\right.
\end{equation}
Moreover, the unit base vectors in Cartesian and spherical coordinates are related through:
\begin{equation}
\arraycolsep=1.5pt%
\begin{array}{lrlllll}
\mathbf{e}_r  = & \sin\theta\cos\phi & \mathbf{e}_x & +\sin\theta \sin\phi & \mathbf{e}_y &+\cos\theta & \mathbf{e}_z\\
\mathbf{e}_\theta  = & \cos\theta\cos\phi & \mathbf{e}_x & +\cos\theta \sin\phi & \mathbf{e}_y &-\sin\theta & \mathbf{e}_z\\
\mathbf{e}_\phi  = & -\sin\phi & \mathbf{e}_x & +\phantom{\sin\theta}\,\cos\phi & \mathbf{e}_y & &
\end{array}
\end{equation}
The inverse relations are:
\begin{equation}
\arraycolsep=1.5pt%
\begin{array}{lllllll}
\mathbf{e}_x  = & \sin\theta\cos\phi & \mathbf{e}_r & +\cos\theta \cos\phi & \mathbf{e}_\theta & -\sin\phi & \mathbf{e}_\phi\\
\mathbf{e}_y  = & \sin\theta\sin\phi & \mathbf{e}_r & +\cos\theta \sin\phi & \mathbf{e}_\theta & +\cos\phi & \mathbf{e}_\phi\\
\mathbf{e}_z  = & \cos\theta & \mathbf{e}_r & -\sin\theta & \mathbf{e}_\theta & &
\end{array}
\end{equation}

\subsection{The spheroid geometry}
\label{sec:geometry}

This code is specific to spheroids, which are described in spherical coordinates as (Fig.~\ref{fig:schematic}):
\begin{align}
  r(\theta) ={}& \frac{ac}{\sqrt{a^2\cos^2\theta + c^2\sin^2\theta}}\\
  \frac{\dd{}r}{\dd\theta}=r_\theta ={}&  \frac{a^2-c^2}{a^2c^2}r(\theta)^3\sin\theta\cos\theta,
\end{align}
where $a$ is the semi-axis length along the $x$- and $y$-axes, and $c$
is the semi-axis length along the $z$-axis, which is the axis of
revolution.

There are two classes of spheroids that may be considered (Fig.~\ref{fig:schematic}). \emph{Oblate} spheroids ($a>c$) are ``smarties''-like (flattened), while \emph{prolate} spheroids ($c>a$) resemble a rugby ball (or a cigar, depending on your inclination). The degenerate case where $a=c$ reduces to a sphere. The aspect ratio, $h$, is defined as the ratio between maximum and minimum distances from the origin:
\begin{align}
  h=\frac{r_\text{max}}{r_\text{min}}=&
  \begin{cases}
    \dfrac{a}{c}&\text{for oblate spheroids,}\\[0.8em]
    \dfrac{c}{a}&\text{for prolate spheroids.}
  \end{cases}
\end{align}
Note that this is different from \cite{1998MishchenkoJQSRT} where the aspect ratio is chosen as $a/c$ and therefore smaller
than unity for prolate spheroids.

Often, spheroids are characterised by their equivalent-volume sphere radius $r_V$, or their equivalent-area sphere radius, $r_A$. The volume of a spheroid is
\begin{align}
  V = \frac{4}{3}\pi{}a^2c
\end{align}
and hence the equivalent-volume radius is
\begin{align}
  r_V = \sqrt[3]{a^2c}.
\end{align}

The surface area of a spheroid is
\begin{align}
  S ={}&
  \begin{cases}
    2\pi{}a^2\Bigl(1+\frac{1-e^2}{e}\tanh^{-1}e\Bigr){}&\text{if
      oblate}\\[0.8em]
    2\pi{}a^2\Bigl(1+\frac{c}{ae}\sin^{-1}e\Bigr){}&\text{if prolate}
  \end{cases}
\end{align}
where $e$ is the eccentricity, which with our definition of the aspect ratio ($h>1$) can be written as $e=\sqrt{h^2-1}/h$ for both types of spheroids.
From this, it is possible to express the equivalent-area sphere radius as
\begin{align}
  r_A ={}&
  \begin{cases}
    a\sqrt{\frac{1}{2} + \frac{1-e^2}{2e}\tanh^{-1}e}{}& \text{if oblate}\\[0.8em]
    a\sqrt{\frac{1}{2}+\frac{c}{2ae}\sin^{-1}e}{}& \text{if prolate}.
  \end{cases}
\end{align}
These values $r_V$ and $r_A$ are provided here for reference, but they are not used explicitly in the code.

\subsection[The \tmatrix/EBCM method]{Principle of the \tmatrix/EBCM method}

The \tmatrix/EBCM method can be viewed as an extension of Mie theory to non-spherical scatterer geometries. In both Mie theory and the \tmatrix\ method, the fields are expanded in terms of vector spherical wavefunctions (VSWFs), as
\begin{align}
\mathbf{E}_\text{inc}=E_0&\sum_{n,m} a_{nm}\mathbf{M}^{(1)}_{nm}\left(k_1\mathbf{r}\right)+b_{nm}\mathbf{N}^{(1)}_{nm}\left(k_1\mathbf{r}\right) \label{EqnEinc}\\
\mathbf{E}_\text{sca}=E_0&\sum_{n,m} p_{nm}\mathbf{M}^{(3)}_{nm}\left(k_1\mathbf{r}\right)+q_{nm}\mathbf{N}^{(3)}_{nm}\left(k_1\mathbf{r}\right) \label{EqnEsca}\\
\mathbf{E}_\text{int}=E_0&\sum_{n,m} c_{nm}\mathbf{M}^{(1)}_{nm}\left(k_2\mathbf{r}\right)+d_{nm}\mathbf{N}^{(1)}_{nm}\left(k_2\mathbf{r}\right) \label{EqnEint}
\end{align}
where for convenience the external field is decomposed into the sum of incident and scattered fields as $\mathbf{E}_\mathrm{out} =
\mathbf{E}_\mathrm{inc} + \mathbf{E}_\mathrm{sca}$. $k_1$ ($k_2$) is the wavevector in the embedding medium (particle), $\mathbf{M}^{(1)}$ and $\mathbf{N}^{(1)}$ are the magnetic and electric regular (finite at the origin) VSWFs, and $\mathbf{M}^{(3)}$ and $\mathbf{N}^{(3)}$ are the irregular magnetic and electric VSWFs that satisfy the radiation condition for outgoing spherical waves. The indices $m$ and $n$ correspond to the projected and total angular momentum, respectively with $|m| \leq n$ and $n=1 \ldots \infty$. The VSWFs definition can be found in Appendix C of Ref.~\cite{2002Mishchenko}.

A unit incident field ($E_0=1$) is assumed everywhere in the code (by linearity, the fields scale proportionally to $E_0$). We also note that all fields here refer to the time-independent complex fields (or phasors), which represent harmonic monochromatic fields $\mathbf{\tilde E}(t)$ of angular frequency $\omega$ using the following convention:
\begin{equation}
\mathbf{\tilde E}(\mathbf{r},t) = \mathrm{Re}\left(\mathbf{E}(\mathbf{r})e^{-i\omega t}\right).
\end{equation}

By linearity of the scattering equations, the expansion coefficients are linearly related and we can define four matrices as follows:
\begin{align}
 \begin{pmatrix}
\mathbf{p}\\\mathbf{q}
\end{pmatrix}  &=-\mathbf{P} \begin{pmatrix}
\mathbf{c}\\\mathbf{d}
\end{pmatrix},&
 \begin{pmatrix}
\mathbf{a}\\\mathbf{b}
\end{pmatrix}  &= \mathbf{Q} \begin{pmatrix}
\mathbf{c}\\\mathbf{d}
\end{pmatrix},\label{Eqnpqab}
\\[1em]
 \begin{pmatrix}
\mathbf{p}\\\mathbf{q}
\end{pmatrix}  &=\phantom{-} \mathbf{T} \begin{pmatrix}
\mathbf{a}\\\mathbf{b}
\end{pmatrix},&
 \begin{pmatrix}
\mathbf{c}\\\mathbf{d}
\end{pmatrix}  &= \mathbf{R} \begin{pmatrix}
\mathbf{a}\\\mathbf{b}
\end{pmatrix},
\label{Eqnpqcd}
\end{align}
where the expansions coefficients are formally grouped in vectors $\mathbf{a}, \mathbf{b}, \mathbf{c}, \mathbf{d}, \mathbf{p}, \mathbf{q}$ with a combined index $p \equiv (n,m)$.

Each of these matrices can be written in block notation as follows, with the block index referring to the type of multipole (electric or magnetic),
\begin{align}
 \mathbf{Q}=
\begin{pmatrix}
\mathbf{Q}^{11}& \mathbf{Q}^{12} \\[1em]
\mathbf{Q}^{21}&\mathbf{Q}^{22}
\end{pmatrix}.
\end{align}
Each block is an infinite square matrix, which is in practice truncated to only include elements acting on multipole orders up to a maximum order $N$. Taking into account $|m|\le n$, each block in the matrix has dimensions $N(N+2)\times N(N+2)$. The most common method to calculate those matrices is the Extended Boundary-Condition Method (EBCM) also called the Null-Field Method, where the matrix elements of $\mathbf{P},\mathbf{Q}$ are obtained as surface integrals on the particle
as derived for example in Ref.~\cite{2002Mishchenko}, Sec.~5.8.

In practice, the expansion coefficients of the incident field ($\mathbf{a},\mathbf{b}$) are known, and the scattered field can be obtained from $\mathbf{T}$, while the internal field results from $\mathbf{R}$. From the above equations, those two matrices can be computed
from $\mathbf{P}$ and $\mathbf{Q}$ as:
\begin{align}
 \mathbf{T} = -\mathbf{PQ}^{-1},\qquad \mathbf{R} = \mathbf{Q}^{-1}.
\end{align}

The matrix $\mathbf{T}$ contains all information about the scatterer. It allows in particular for analytical averaging over all orientations \cite{1991MishchenkoJOSAA, 1992KhlebtsovAO,1998MishchenkoJQSRT, 1993MishchenkoAO, 1996MackowskiJOSAA} or solving multiple scattering problems by an ensemble of particles \cite{1996MackowskiJOSAA, 1973PetersonPRD, 1993MishchenkoAO}.


\subsection{Additional simplifications for spheroids}

For particles with symmetry of revolution, such as spheroids, expansion coefficients
with different $m$ values are entirely decoupled, and one can therefore solve the problem for each value of $m$, where $m$ can be viewed as a fixed parameter (which will be implicit in most of our notations).
This means that each large $2N(N+2) \times 2N(N+2)$ matrix can be decoupled into $2N+1$ independent matrices with $m=-N\dots N$, each of size $2(N-m+1)\times 2(N-m+1)$ (or $2N\times 2N$ for $m=0$).
Moreover, we have:
\begin{align}
T^{11}_{n,k|-m} &= \phantom{-}T^{11}_{n,k|m},& T^{12}_{n,k|-m} &= -T^{21}_{n,k|m},  \nonumber\\[0.2cm]
T^{21}_{n,k|-m} &= -T^{12}_{n,k|m},& T^{22}_{n,k|-m} &= \phantom{-}T^{22}_{n,k|m} .
\end{align}
and therefore only $m\ge 0$ values need to be considered in the calculation of $\mathbf{T}$.

Furthermore, the surface integrals reduce to line integrals, for which we have recently proposed a number of simplified expressions \cite{2011SomervilleOL}.

Reflection symmetry with respect to the equatorial plane also results in a number of additional simplifications (see Sec.~5.2.2 of Ref.~\cite{2002Mishchenko} and Sec.~2.3 of Ref.~\cite{JQSRT2013}). Half of the matrix entries are zero because of the symmetry in changing $\theta\rightarrow\pi-\theta$ and the other integrals are simply twice the integrals evaluated over the half-range $0$ to $\pi/2$. Explicitly, we have
\begin{alignat}{3}
  P^{11}_{nk} &= P^{22}_{nk} &= Q^{11}_{nk} &= Q^{22}_{nk} &= 0 & \quad\text{if~}n+k\text{~odd}\nonumber, \\
  P^{12}_{nk} &= P^{21}_{nk} &= Q^{12}_{nk} &= Q^{21}_{nk} &= 0 & \quad\text{if~}n+k\text{~even},
\end{alignat}
and identical relations for $\mathbf{T}$ and $\mathbf{R}$. From the point of view of numerical implementation, it means that only half the elements need to be computed and stored. More importantly, it also implies that we can rewrite Eqs.~\ref{Eqnpqab}-\ref{Eqnpqcd} as two independent sets of equations \cite{JQSRT2013}.
Explicitly, we define
\begin{align}
\mathbf{a}_e=  \begin{pmatrix}
a_2\\a_4\\\vdots
\end{pmatrix},
\mathbf{b}_o=  \begin{pmatrix}
b_1\\b_3\\\vdots
\end{pmatrix},
\mathbf{a}_o=  \begin{pmatrix}
a_1\\a_3\\\vdots
\end{pmatrix},
\mathbf{b}_e=  \begin{pmatrix}
b_2\\b_4\\\vdots
\end{pmatrix},
\end{align}
and similarly for $\mathbf{c}$, $\mathbf{d}$, $\mathbf{p}$, $\mathbf{q}$. We also define the matrices $\mathbf{Q}^{eo}$ and $\mathbf{Q}^{oe}$
from $\mathbf{Q}$ as:
\begin{align}
\mathbf{Q}^{eo}=\begin{pmatrix}
\mathbf{Q}^{11}_{ee} & \mathbf{Q}^{12}_{eo} \\[0.4cm]
\mathbf{Q}^{21}_{oe} & \mathbf{Q}^{22}_{oo}
\end{pmatrix},\qquad
\mathbf{Q}^{oe}=\begin{pmatrix}
\mathbf{Q}^{11}_{oo} & \mathbf{Q}^{12}_{oe} \\[0.4cm]
\mathbf{Q}^{21}_{eo} & \mathbf{Q}^{22}_{ee}
\end{pmatrix},
\label{EqnBlockEOOE}
\end{align}
where $Q^{12}_{eo}$ denotes the submatrix of $Q^{12}$ with even row indices
and odd column indices, and similarly for the others. One can see that $\mathbf{Q}^{eo}$ and $\mathbf{Q}^{oe}$ contain all the non-zero elements of $\mathbf{Q}$ and exclude all the elements that must be zero by reflection symmetry, so this is an equivalent description of the $Q$-matrix.

The equations relating the expansion coefficients then decouple into two sets of independent equations, for example
\begin{equation}
\begin{pmatrix} \mathbf{a}_e\\ \mathbf{b}_o\end{pmatrix}=\mathbf{Q}^{eo}
\begin{pmatrix} \mathbf{c}_e\\ \mathbf{d}_o\end{pmatrix},\qquad
\begin{pmatrix} \mathbf{a}_o\\ \mathbf{b}_e\end{pmatrix}=\mathbf{Q}^{oe}
\begin{pmatrix} \mathbf{c}_o\\ \mathbf{d}_e\end{pmatrix},
\end{equation}
and similar expressions deduced from Eqs.~\ref{Eqnpqab}--\ref{Eqnpqcd} for $\mathbf{P}$, $\mathbf{T}$, and $\mathbf{R}$. As a result, the problem of finding the $2(N-m+1)\times 2(N-m+1)$ $T$- (or $R$-) matrix up to multipole order $N$ reduces to finding the two decoupled $T$-matrices $\mathbf{T}^{eo}$ and $\mathbf{T}^{oe}$, each of size $(N-m+1)\times (N-m+1)$, namely:
\begin{align}
\mathbf{T}^{eo} = -\mathbf{P}^{eo}\left(\mathbf{Q}^{eo}\right)^{-1},\qquad
\mathbf{T}^{oe} = -\mathbf{P}^{oe}\left(\mathbf{Q}^{oe}\right)^{-1}.
\end{align}

These symmetries and the definitions of this section are used in the code to compute and store the matrices.

\subsection{Angular functions}

The \tmatrix\ integrals and many of the physical properties are expressed in terms of angular ($\theta$-dependent) functions, which are derived from the associated Legendre functions $P_n^m(x)$. We here summarize the most important definitions. The associated Legendre functions may be written in terms of the Legendre polynomials as (for $m\ge0$)
\begin{align}
  P^m_n(x) = (-1)^m\left(1-x^2\right)^{m/2}\frac{\dd^m}{\dd{}x^m}P_n(x)\label{eq:assoc_legendre}
\end{align}
where the polynomial is given by the expression
\begin{align}
  P_n(x) = \frac{1}{2^nn!}\frac{\dd^n}{\dd{}x^n}\left(x^2-1\right)^n.
\end{align}
The factor $(-1)^m$ in the definition of Eq.~(\ref{eq:assoc_legendre}) is known as the Condon-Shortley phase. In the case of negative $m$, the expression for the associated Legendre function is
\begin{align}
  P^{-m}_n(x) = (-1)^m\frac{(n-m)!}{(n+m)!}P^{m}_n(x).
\end{align}

Following Ref.~\cite{2002Mishchenko}, we do not use the associated Legendre functions directly, but rather some functions obtained from them, which have more favorable numerical properties. We notably use a special case of the Wigner $d$-functions,
\begin{align}
  d_{nm}(\theta) \equiv d^n_{0m}(\theta)=(-1)^m\sqrt{\frac{(n-m)!}{(n+m)!}}P^m_n(\cos\theta)
\end{align}
where we make use of the simpler $d_{nm}(\theta)$ notation.
We will also use the functions $\pi_{nm}(\theta)$ and
$\tau_{nm}(\theta)$, derived from them as (Eqs.~5.16 and 5.17 of \cite{2002Mishchenko}):
\begin{align}
  \pi_{nm}(\theta) {}&=  \frac{md_{nm}(\theta)}{\sin\theta},\nonumber\\
 \tau_{nm}(\theta) {}&=  \frac{\dd}{\dd{}\theta}d_{nm}(\theta).
\end{align}

The function $\pi_{nm}(\theta)$ is generated for $m>0$ using the recursion relation (derived from Eq.~B.22 in \cite{2002Mishchenko}, see also \cite{2000MishchenkoAO}):
\begin{align}
  \pi_{n,m}(\theta) &= \frac{1}{\sqrt{n^2-m^2}}\left((2n-1)\cos\theta~\pi_{n-1,m}(\theta)\right)\nonumber\\[0.2cm]
	&-\sqrt{(n-1)^2-m^2}\pi_{n-2,m}(\theta),
\end{align}
applied for $n\geq m+1$ with the initial conditions
\begin{align}
  \pi_{m-1,m}(\theta) {}&= 0 \nonumber\\
\pi_{m,m}(\theta) {}&= mA_m(\sin\theta)^{m-1}
\intertext{with $A_m$ defined recursively as}
A_0{}&=1 \nonumber\\
A_{m+1}{}&=A_m\sqrt{\frac{2m+1}{2(m+1)}}.
\end{align}
$\tau_{nm}$ is then calculated as
\begin{align}
  \tau_{nm}(\theta) =
  \frac{-1}{m}\sqrt{n^2-m^2}\pi_{n-1,m}(\theta)+\frac{n}{m}\cos\theta~\pi_{nm}(\theta).
\end{align}
For $m<0$, the following relations are used:
\begin{align}
  \pi_{n,-m}(\theta) {}&= (-1)^{m+1}\pi_{nm}(\theta), \nonumber\\[0.2cm]
  \tau_{n,-m}(\theta) {}&= (-1)^{m}\tau_{nm}(\theta).
\end{align}
Finally, for $m=0$, we have
\begin{align}
  \pi_{n,0}(\theta) {}&= 0 \nonumber\\
  \tau_{n,0}(\theta) {}&= - \sin\theta~ P'_n(\cos\theta).
\end{align}
When $d_{nm}(\theta)$ is needed, it is calculated as:
\begin{align}
\begin{cases}
d_{n,m} (\theta) = \frac{1}{m}\sin\theta~ \pi_{n,m}(\theta) & \mathrm{if~}m\neq 0\\[0.2cm]
d_{n,0} (\theta) = P_n(\cos\theta) & \mathrm{if~}m= 0
\end{cases}
\end{align}
In those latter expressions, $d_{n,0}$ and $\tau_{n,0}$ are obtained by standard recursion for the Legendre polynomials and its derivatives. For $n \geq 1$:
\begin{align}
d_{n,0} &=\tfrac{2n-1}{n}\cos\theta ~d_{n-1,0}-\tfrac{n-1}{n}d_{n-2,0}\nonumber\\[0.2cm]
\tau_{n,0} &=\cos\theta ~\tau_{n-1,0}-n\sin\theta ~ \tau_{n-2,0}\nonumber\\[0.2cm]
 d_{-1,0}&=0,\quad d_{0,0}=1 \nonumber\\
 \tau_{-1,0}&=0,\quad\tau_{0,0}=-\sin\theta.
\end{align}

The function \texttt{vshPinmTaunm} computes the required angular functions using the above formulas, which are numerically stable and efficient \cite{2002Mishchenko}.

\subsection{Integral quadratures}

All \tmatrix\ integrals can be written as integrals over the variable $\cos(\theta)$. The integrals are numerically computed using a standard Gauss-Legendre quadrature scheme with $N_\theta$ points,
\begin{equation}
\int_0^\pi f(\theta)\sin\theta\, \dd\theta =  \int_{-1}^1 f(\theta)\,\dd(\cos\theta) \approx \sum_{p=1}^{N_\theta} w_p f(\theta_p),
\end{equation}
where $\theta_p$ and $w_p$ are the nodes and weights of the quadrature. The same procedure is also used for calculating surface-averaged field properties, but may require a different number of integration points.

The function \texttt{auxInitLegendreQuad} calculates these nodes and weights for any number $N_\theta$ and uses the algorithm developed by Greg von Winckel available from the \MATLAB\ Central website \cite{Winckel:aa} (where it is called \texttt{lgwt.m}). For convenience the file \texttt{Utils/quadTable.mat} stores pre-calculated nodes and weights by steps of 5 from 50 up to 2000, which can reduce the calculation time.

For spheroids, the \tmatrix\ integrals can be reduced to a half-interval by symmetry, so the nodes and weights are computed for quadrature order $2N_\theta$ and only the positive nodes $\theta_p>0$ are used (giving $N_\theta$ quadrature points).

We note that alternative quadrature schemes could easily be used, and may perform better for these types of integrands (requiring fewer function evaluations). Unfortunately, in order to make the best use of vectorised calculations, paramount for efficient \MATLAB\ code, the implementation of adaptive quadrature (with internal accuracy estimate) appears challenging and would require an important refactoring of those functions performing numerical integrations.

\subsection{Computation of the $\mathbf{P}$ and $\mathbf{Q}$ matrices}

The formulas used for the computation of the integrals of the $\mathbf{P}$ and $\mathbf{Q}$ matrices
are given in Sec.~2.2 of Ref.~\cite{JQSRT2013}.
Explicitly, we use the following equations from Ref.~\cite{JQSRT2013}:
\begin{align}
\begin{cases}
\mathbf{P}^{12},~\mathbf{Q}^{12}: & \text{Eqs.~11,15} \\
\mathbf{P}^{21},~\mathbf{Q}^{21}: & \text{Eqs.~12,16} \\
\mathbf{P}^{11},~\mathbf{Q}^{11}: & \text{Eqs.~17,18} \\
\mathbf{P}^{22},~\mathbf{Q}^{22}: & \text{Eqs.~19--22}
\end{cases}
\end{align}
The diagonal terms are treated separately and we use:
\begin{align}
\begin{cases}
P_{nn}^{11},~{Q}_{nn}^{11}: & \text{Eqs.~23,25,65} \\
P_{nn}^{22},~{Q}_{nn}^{22}: & \text{Eqs.~24,26,27} \\
\end{cases}
\end{align}

The algorithm used to avoid numerical cancellations was described in detail in \cite{JQSRT2013} and summarized in Sec.~4.4 of \cite{JQSRT2013}. All the technical details of the implementation can be found in Ref.~\cite{JQSRT2013}, in particular in the Appendix. Comments in the code also explain the most important steps, using the same notation and referring to equations and sections of Ref.~\cite{JQSRT2013}.

The function \texttt{sphCalculatePQ} handles all those calculations and returns the two matrices.

The functions  \texttt{sphGetModifiedBesselProducts}, \texttt{sphGetXiPsi}, and \texttt{sphGetFpovx} are used specifically to implement the new algorithm.

One important parameter of the new algorithm is the number of multipoles, $N_B$, used to estimate the modified Bessel products. For large size parameters, it may be necessary to use $N_B>N_Q$ to obtain accurate results. Whether this precaution is necessary can be easily checked before carrying out the bulk of the calculations. The function \texttt{sphEstimateNB} can be called to provide such an estimate for $N_B$. It calculates the modified Bessel products ($F^+/x$) for the maximum size parameter and the smallest and largest $s$ (if $\lambda$-dependent) for increasing $N_B$ until all results up to $n=N_Q$ have converged (within a specified relative accuracy, the default value is $10^{-13}$).

\subsection{Matrix inversion for $\mathbf{T}$ and $\mathbf{R}$ matrices}
\label{Tcomp}

The inversion of the linear systems for $\mathbf{T}$ and $\mathbf{R}$ is performed  using block inversion as detailed in Sec.~4.5 of \cite{JQSRT2013}. Specifically, the inversion is carried out with the following steps (Eq.~70 of \cite{JQSRT2013}),
\begin{align}
\mathbf{F}_1 ={}& \left(\mathbf{Q}^{11}\right)^{-1}, \nonumber\\
\mathbf{G}_1 ={}& \mathbf{P}^{11}\mathbf{F}_1,\quad
\mathbf{G}_3 ={} \mathbf{P}^{21}\mathbf{F}_1,\quad
\mathbf{G}_5 ={} \mathbf{Q}^{21}\mathbf{F}_1. \nonumber\\[0.2cm]
\mathbf{F}_2 ={}& \left[\mathbf{Q}^{22} -
\mathbf{G}_5\mathbf{Q}^{12}\right]^{-1} \nonumber,\\
\mathbf{G}_2 ={}& \mathbf{P}^{22}\mathbf{F}_2,\quad
\mathbf{G}_4 ={} \mathbf{P}^{12}\mathbf{F}_2,\quad
\mathbf{G}_6 ={} \mathbf{Q}^{12}\mathbf{F}_2. \nonumber\\[0.2cm]
\mathbf{T}^{12} ={}& \mathbf{G}_1\mathbf{G}_6-\mathbf{G}_4, \quad
\mathbf{T}^{22} ={} \mathbf{G}_3\mathbf{G}_6-\mathbf{G}_2, \nonumber\\
\mathbf{T}^{11} ={}& \mathbf{G}_1 - \mathbf{T}^{12}\mathbf{G}_5, \quad
\mathbf{T}^{21} ={} \mathbf{G}_3 - \mathbf{T}^{22}\mathbf{G}_5.
\end{align}
This is carried out separately for $\mathbf{T}^{eo}$ and
$\mathbf{T}^{oe}$.

Two matrix inversions are needed in those steps (to compute $\mathbf{F}_1$ and $\mathbf{F}_2$). Because of the often near-singular nature
of the matrices, the choice of the inversion algorithm can have dramatic consequences on the numerical stability of the calculations.
A number of options have been proposed and studied in the literature.
In \cite{1997WielaardAO}, a method based on a LU factorization with partial row pivoting (equivalent to ${\tt A/B}$ in \MATLAB\ to get $\mathbf{A}\mathbf{B}^{-1}$)
was proposed. In \cite{JQSRT2013}, we observed that {\tt (B.\textquotesingle\textbackslash A.\textquotesingle).\textquotesingle} appeared to
be more numerically stable. This amounts to a LU factorization with partial {\it column} pivoting (as opposed to row pivoting as suggested in \cite{1997WielaardAO}). Although not explicitly stated as such, we believe this is equivalent to the improved
algorithm proposed in \cite{2005MorozAO} and based on Gaussian elimination with back-substitution.

In \CODENAME, we implement the inversion algorithm explicitly to avoid using the ${\tt \textbackslash}$ operator, which
has a different behavior in \MATLAB\ and Octave for near-singular matrices. The steps are as follows. The function ${\tt lu}$ is called on
the transpose of the matrix, $\mathbf{B}^\mathbf{T}$, to enforce column pivoting instead of rows, i.e. we obtain lower and upper triangular matrices
$\mathbf{L}$ and $\mathbf{U}$ and a permutation matrix $\mathbf{P}$ such that
\begin{align}
\mathbf{L} \mathbf{U} = \mathbf{P} \mathbf{B}^\mathbf{T}.
\end{align}
The solution of $\mathbf{X} \mathbf{B} = \mathbf{A}$ is then obtained by successively solving the following two triangular linear systems
and transposing the result, i.e.
\begin{align}
\mathbf{L} \mathbf{Z} &= \mathbf{P} \mathbf{A}^\mathbf{T}\\
\mathbf{U} \mathbf{Y} &= \mathbf{Z}\\
\mathbf{X} &= \mathbf{Y}^\mathbf{T}
\end{align}
$\mathbf{F}_1$ and $\mathbf{F}_2$ are calculated with this algorithm by setting $\mathbf{A}=\mathbf{I}$.
Note that with this algorithm, we have not noticed any difference in accuracy when calculating $\mathbf{T}$
directly from solving $\mathbf{T}\mathbf{Q}=-\mathbf{P}$ as opposed to calculating $\mathbf{R}$ first
from $\mathbf{R}\mathbf{Q}=\mathbf{I}$ and then deducing $\mathbf{T}$ from $\mathbf{T}=-\mathbf{P}\mathbf{R}$.

The function \texttt{rvhGetTRfromPQ} calculates $\mathbf{T}$ (and optionally $\mathbf{R}$).

Note that, as explained in detail in Ref.~\cite{JQSRT2015}, the elements of the \tmatrix\ are not accurate up to multipole $n=N_Q$ even when $\mathbf{P}$ and $\mathbf{Q}$ are.
If an accurate \tmatrix\ up to multipole $N$ is required, it is therefore necessary to
calculate  $\mathbf{P}$ and $\mathbf{Q}$ with $N_Q=N+\Delta$ multipoles, and then truncate the obtained \tmatrix\ down to $N$ multipoles (see \cite{JQSRT2015} for full details). In such cases, the function \texttt{sphEstimateDelta} can be used to estimate $\Delta$ and the function \texttt{rvhTruncateMatrices} is then used to truncate $T$ down to $N$ multipoles.

In principle, the \tmatrix\ should satisfy general symmetry relations arising from optical reciprocity \cite{2002Mishchenko,JQSRT2015}, namely:
\begin{align}
T_{nk}^{11} = \phantom{-}T_{kn}^{11},\quad & T_{nk}^{21} = -T_{kn}^{12},\nonumber\\
T_{nk}^{12} = -T_{kn}^{21},\quad & T_{nk}^{22} = \phantom{-}T_{kn}^{22}.
\end{align}
It was suggested in \cite{2013VolkovJQSRT,JQSRT2015} that the upper triangular part of the \tmatrix\ is more accurate in challenging cases than the lower triangular part. Using the function \texttt{rvhGetSymmetricMat}, one can use these symmetry relations to deduce the lower parts from the upper parts. This can slightly increase
the range of validity of the method.

As pointed out in \cite{JQSRT2015}, these precautions are not necessary in many cases, and it is sufficient to check that the desired physical properties have converged (see convergence tests in Sec.~\ref{sec:convergence}).

\subsection{Orientation-averaged properties}

One of the advantages of the \tmatrix\ formalism is that the optical properties for any orientation can in principle be derived from a single computation of the scatterer \tmatrix. In particular, once the \tmatrix\ has been calculated, it is possible to calculate analytically the optical properties of a (non-interacting) collection of randomly oriented scatterers. Such orientation-averaged far-field cross-sections are evaluated as detailed in Ref.~\cite{2002Mishchenko}. We have in particular (Eqs. 5.107 and 5.141 of \cite{2002Mishchenko}):
\begin{align}
\langle{}C_\text{ext}\rangle ={}&
\frac{-2\pi}{k^2_1}\sum_{n,m}\mathrm{Re}\left(T^{11}_{nn|m}+T^{22}_{nn|m}\right), \nonumber\\
=& \frac{-2\pi}{k^2_1}\sum_{\substack{ n=1\dots\infty \\[0.05cm] m=0\dots n}} (2-\delta_{m,0})\mathrm{Re}\left(T^{11}_{nn|m}+T^{22}_{nn|m}\right)
\end{align}
\begin{align}
& \langle{}C_\text{sca}\rangle ={}
\frac{2\pi}{k_1^2}\sum_{\substack{ n=1\dots\infty \\[0.05cm] k=1\dots\infty \\[0.05cm]  m=0\dots\min(n,k)}} (2-\delta_{m,0}) \times\nonumber \\
&\left(\left|T^{11}_{nk|m}\right|^2+\left|T^{12}_{nk|m}\right|^2+\left|T^{21}_{nk|m}\right|^2+\left|T^{22}_{nk|m}\right|^2\right)\label{eq:Csca_oa}
\end{align}
\begin{align}
\langle{}C_\text{abs}\rangle
={}&\langle{}C_\text{ext}\rangle-\langle{}C_\text{sca}\rangle.
\end{align}

The function \texttt{rvhGetAverageCrossSections} calculates those cross-sections from a previously-obtained \tmatrix.

\subsection[Scattering matrix]{Scattering matrix for random orientation}

The \tmatrix\ formalism can also be used to efficiently and accurately compute the scattering matrix for randomly-oriented scatterers. The full details of such calculations are described in Sec.~5.5 of Ref.~\cite{2002Mishchenko} and the corresponding algorithm has been implemented in standard \tmatrix\ codes~\cite{1998MishchenkoJQSRT}. For convenience, we here provide a function \texttt{pstScatteringMatrixOA} to calculate this scattering matrix and output the results in the same format as in Ref.~\cite{1998MishchenkoJQSRT}. This function (and the subroutines it uses) are a direct port of those \FORTRAN\ routines into \MATLAB\ and are here provided for convenience with permission from M.I.~Mishchenko. Because no attempt was made to optimize them for \MATLAB, they are much slower than the corresponding \FORTRAN\ routines. For any intensive scattering matrix calculations, it is therefore recommended to export the \tmatrix\ obtained from \MATLAB\ and run the calculations in \FORTRAN\ using the code of Ref.~\cite{1998MishchenkoJQSRT}.

\subsection{Incident field}

For scatterers with a fixed orientation, one first needs to define the incident field through its corresponding expansion coefficients $a_{nm}$ and $b_{nm}$ (Eq.~\ref{EqnEinc}). Only incident plane waves with linear polarisation are currently implemented in the code. For a general incident plane wave, those are given in Eqs.~(C.56-C59) of Ref.~\cite{2002Mishchenko}. Explicitly, the field is:
\begin{align}
\mathbf{E}(\mathbf{r}) = \mathbf{E}_0 \exp\left(i\mathbf{\mathbf{k}_1}\cdot \mathbf{r}\right)
\end{align}
and we define the incident $k$-vector direction with its two angles from spherical coordinates $\theta_p$, $\phi_p$, i.e.:
\begin{align}
\mathbf{k}_1 & =  k_1\mathbf{e}_{r_p} \nonumber\\[0.2cm]
& = k_1\left(\sin\theta_p\cos\phi_p \mathbf{e}_x + \sin\theta_p\sin\phi_p \mathbf{e}_y +\cos\theta_p \mathbf{e}_z \right).
\end{align}
The incident field polarisation, which must be perpendicular to $\mathbf{k}_1$ is then defined by one angle $\alpha_p$ as:
\begin{align}
\mathbf{E}_0 = E_0 &\left(\cos\alpha_p \mathbf{e}_{\theta_p} + \sin\alpha_p\mathbf{e}_{\phi_p}\right) \nonumber\\[0.2cm]
= E_0 & \left[ \left(\cos\alpha_p \cos\theta_p\cos\phi_p - \sin\alpha_p\sin\phi_p\right)\mathbf{e}_{x} \right. \nonumber\\
& +\left(\cos\alpha_p \cos\theta_p\sin\phi_p + \sin\alpha_p\cos\phi_p\right)\mathbf{e}_{y} \nonumber\\
& \left. - \cos\alpha_p\sin\theta_p \mathbf{e}_z\right]
\end{align}

With those defined, the expansion coefficients are then obtained from:
\begin{align}
a_{nm} &= \bar{d}_{nm}\left[i\cos\alpha_p \pi_{nm}(\theta_p) + \sin\alpha_p\tau_{nm}(\theta_p)\right] \nonumber \\[0.2cm]
b_{nm} &= \bar{d}_{nm}\left[i\cos\alpha_p \tau_{nm}(\theta_p) +\sin\alpha_p\pi_{nm}(\theta_p)\right]
\end{align}
where
\begin{align}
\bar{d}_{nm} = (-1)^{m+1} \exp(-im\phi_p) \times i^n \sqrt{\frac{4\pi (2n+1)}{n(n+1)}}.
\end{align}
Note that if the incident field is incident along the $z$ direction, then only
$|m|=1$ terms are non-zero.\\

Here are a few examples of common configurations:
\begin{align}
\mathrm{KzEx}:&& \theta_p &= 0,& \phi_p &= 0,& \alpha_p&=0\\
\mathrm{KzEy}:&& \theta_p &= 0,& \phi_p &= 0,& \alpha_p&=\pi/2\\
\mathrm{KxEz}:&& \theta_p &= \pi/2,& \phi_p &= 0,& \alpha_p&=\pi\\
\mathrm{KxEy}:&& \theta_p &= \pi/2,& \phi_p &= 0,& \alpha_p&=\pi/2
\end{align}
The function \texttt{vshMakeIncidentParameters} can be used to define these parameters, and \texttt{vshGetIncidentCoefficients} to get the incident field coefficients. We note that these definitions were chosen for linear polarisation, but elliptic polarisation could be easily accommodated by amending the function \texttt{vshGetIncidentCoefficients}.

\subsection[Scattered and internal fields]{Expansion coefficients of scattered and internal fields and fixed-orientation cross-sections}

Once the incident field expansion coefficients are defined, it is straightforward to obtain those of the scattered and internal fields from Eq.~\ref{Eqnpqcd}. The function \texttt{rvhGetExpansionCoefficients} will carry out this task. The internal fields coefficients are only computed if the matrix $\mathbf{R}$ was calculated.

Once the expansion coefficients of the scattered field are known, the extinction, scattering, and absorption cross-sections are simply obtained from similar expressions as for standard Mie theory (Eq.~5.18 of \cite{2002Mishchenko}):
\begin{align}
 C_{\text{sca}} ={}&
\frac{1}{k_1^2}\sum_{n,m}\left(\left|c_{nm}\right|^2+\left|d_{nm}\right|^2\right)\\
C_\text{ext} ={}&
\frac{-1}{k^2_1}\sum_{n,m}\left(c^*_{nm}a_{nm} +
d^*_{nm}b_{nm}\right)\\ C_\text{abs} ={}&
C_\text{ext}-C_\text{sca}.
\end{align}

\subsection{Surface fields}

\tmatrix\ calculations have been mostly applied to far-field properties but for many applications in plasmonics, nanophotonics, optical forces, etc., the near-field properties are also needed.

The applicability of the \tmatrix\ method to near-field calculations is still
debated; a particular point of concern is to avoid reliance on the Rayleigh hypothesis, which is generally not valid. This implies that the scattered field expansion (Eq.~\ref{EqnEsca}) is no longer valid for fields near the scatterer surface (but it can be shown that it is valid at least outside the circumscribing sphere of the scatterer \cite{2002Mishchenko}).

To circumvent this limitation, we here use an alternative approach relying on the internal field expansion (Eq.~\ref{EqnEint}), which remains valid everywhere at the surface (at least in the case of spheroids). This expansion allows us to calculate the internal field everywhere on the surface (but inside) of the particle, $\bar{\mathbf{E}}^\mathrm{in}$. In order to calculate fields $\bar{\mathbf{E}}^\mathrm{out}$ immediately outside the surface, we apply the standard boundary conditions:
\begin{align}
  \left(\bar{\mathbf{E}}^\mathrm{int} - \bar{\mathbf{E}}^\mathrm{out}\right)\times\mathbf{n}= \mathbf{0} \nonumber\\
  \left(\varepsilon_\mathrm{in}\bar{\mathbf{E}}^\mathrm{int} - \varepsilon_\mathrm{out}\bar{\mathbf{E}}^\mathrm{out}\right)\cdot\mathbf{n}= 0
\end{align}
where the normal is
\begin{align}
\mathbf{n} &= n_r\mathbf{e}_r + n_\theta \mathbf{e}_\theta \nonumber\\
& =\frac{r}{\sqrt{r^2+r_\theta^2}}\mathbf{e}_r - \frac{r_\theta}{\sqrt{r^2+r_\theta^2}}\mathbf{e}_\theta.
\end{align}
Explicitly, we have (using $s^2=\varepsilon_\mathrm{2}/\varepsilon_\mathrm{1}$)
\begin{align}
  \bar{{E}}^\mathrm{out}_r &= \left[1+(s^2-1)n_r^2\right] \bar{{E}}^\mathrm{in}_r
	+ \left[s^2-1\right] n_r n_\theta \bar{{E}}^\mathrm{in}_\theta \nonumber \\[0.2cm]
  \bar{{E}}^\mathrm{out}_\theta &= \left[s^2-1\right] n_r n_\theta \bar{{E}}^\mathrm{in}_r
	+ \left[1+(s^2-1)n_\theta^2\right] \bar{{E}}^\mathrm{in}_\theta \nonumber \\[0.2cm]
  \bar{{E}}^\mathrm{out}_\phi &= \bar{{E}}^\mathrm{in}_\phi
\end{align}

The function \texttt{pstSurfaceFields} uses this method to calculate the surface electric field along with a number of surface-averaged properties relevant to plasmonics and other near-field applications. Note that the $\phi$-dependence of all quantities is relatively simple, since they can all be expressed as:
\begin{align}
A(r,\theta,\phi) = \sum_{m=-N}^{m=+N} A_m(r,\theta)\exp(im\phi).
\end{align}
Our code therefore calculates the $2N+1$ variables $A_m(r,\theta)$ and the $\phi$-dependence can then be trivially re-introduced.

\subsection{Near fields}

This method of calculating surface fields, however, cannot be used to obtain the near-field except exactly at the surface. For points sufficiently far from the particle, the scattered field may be obtained from Eq.~(\ref{EqnEsca}). However, for particles deviating from a sphere, the series may not converge for points close to the particle (failure of the Rayleigh hypothesis). To test whether a point converges, an indicative test is to increase the number of multipoles considered, and confirm that the calculated field converges to some value. If it fails to converge, either insufficient multipoles were considered, or the point is in a region where convergence will never be obtained.
We have recently studied these aspects and developed an alternative method of computing near-fields, which will be discussed in detail elsewhere \cite{2016RH}.
We only here give a brief overview of the method, which is implemented in the function \texttt{pstGetNearField}.

As originally suggested in Ref. \cite{2010DoicuJQSRT}, we make use of the surface integral equation (Eq. 5.168 of \cite{2002Mishchenko}), which expresses the scattered field in terms of the surface fields,
\begin{align}
\mathbf{E}_\mathrm{sca}(\mathbf{r}')=\int_S dS\left\{i\omega \mu_0\left[\mathbf{n} \times \mathbf{H}(\mathbf{r})\right]\cdot\overleftrightarrow{G}(\mathbf{r},\mathbf{r}') \right.\nonumber\\
+ \left.\left[\mathbf{n} \times \mathbf{E}(\mathbf{r})\right]\cdot\left[\bm{\nabla} \times \overleftrightarrow{G}(\mathbf{r},\mathbf{r}')\right]\right\},
\end{align}
where $\overleftrightarrow{G}$ is the free-space Green's function. The surface fields $\mathbf{E}, \mathbf{H}$ can be calculated accurately as described earlier and the integral is then performed using a double quadrature on $\theta$ and $\phi$ (with the same number of nodes for simplicity). As a result, this method is slower than the other methods for calculating the scattered field (where applicable), but will exhibit much better convergence behaviour for points near the particle \cite{2016RH}. An example of its use is given in {\tt ScriptTutorial}.

\section{Additional implementation details}
\label{sec:implementation}
\setlength\parindent{0pt}

\subsection[Naming conventions and organization]{File naming conventions and organization}

The first three letters of each function are used to classify functions depending on their roles:
\begin{itemize}
\item
\texttt{slv}: High-level functions solving a specific class of problems.
\item
\texttt{pst}: High-level functions used for post-processing.
\item
\texttt{vsh}: Mostly low-level functions handling the calculations of quantities related to vector spherical wavefunctions.
\item
\texttt{rvh}: \tmatrix\ related functions specific to particles with mirror-reflection symmetry.
\item
\texttt{sph}: \tmatrix\ related functions specific to spheroidal particles.
\item
\texttt{aux}: Auxiliary functions (low-level).
\end{itemize}

Each m-file is also located in a specific folder with the following classification,

\begin{itemize}
  \item
  \texttt{High-Level}: \tmatrix\ related functions most likely to be called directly by the user.
  \item
  \texttt{Low-Level}: \tmatrix\ related functions used by the high-level functions.
  \item
  \texttt{Materials}: Includes dielectric functions of gold (Au) and silver (Ag) for plasmonics applications, as well as an example dielectric function interpolated from tabulated values for silicon (Si).
  \item
  \texttt{Post-processing}: Functions to calculate optical properties.
\item
\texttt{Scripts}: Contains example and tutorial scripts.
\item
\texttt{Solve}: Contains the \texttt{slv} functions used to solve a specific class of problems.
\item
\texttt{Utils}: Includes miscellaneous utility functions, notably to export the \tmatrix, test whether the code is running in Octave or \MATLAB, or generate quadrature nodes and weights.
\end{itemize}

\subsection{Storage of matrices}

All matrices are stored in cell arrays, such as \texttt{CstPQa} of dimensions $1\times M$, where $M$ is the number of $m$ elements in \texttt{absmvec}. As a result, \texttt{CstPQa\{j\}} corresponds to \texttt{m=absmvec(j)}. If all $m$ are computed, then \texttt{absmvec=0:N} and the matrices for a given $m$
are stored in \texttt{CstPQa\{m+1\}}.

Each element \texttt{CstPQa\{j\}} contains a structure \texttt{stPQ} describing the matrices for the corresponding $m$, stored in the following fields:
\begin{itemize}
\item
\texttt{stPQ.CsMatList}: cell array of strings listing the name of the matrices. For example \texttt{stPQ.CsMatList = \{\textquotesingle st4MP\textquotesingle,\textquotesingle st4MQ\textquotesingle\}}.
\item
For each string in this list, two fields are included with \texttt{eo} and \texttt{oe} appended at the end. For example \texttt{stPQ.st4MPeo} and \texttt{stPQ.st4MPoe}. These \texttt{st4M} structures contain the matrix in a form described below, which avoids storing the many zeros that are imposed by the reflection symmetry.
\end{itemize}

For a given maximum number of multipoles $N$, we have $m\leq n\leq N$ and the matrices are square matrices of dimension $N+1-m$ (or $N$ for $m=0$) since only elements $M_{nk}$ with $n,k\geq m$ (or $\geq 1$ for $m=0$) are needed. Therefore for a given $m$, $M(i,j)$ corresponds to $M_{n=i+m-1,k=j+m-1}$ or $M_{n=i+m,k=j+m}$ for $m=0$. Using the reflection symmetry, the matrix $\mathbf{M}$ is moreover written in block $oe-eo$ notation as in Eq.~\ref{EqnBlockEOOE}. In this notation, all the obvious zeros have been removed, and only the relevant $n,k$ pairs are included.

The four blocks are given in a \texttt{st4Meo} or \texttt{st4Moe} structure, which contains the following fields:
\begin{itemize}
\item
\texttt{st4Meo.m}: the $m$-value the matrix corresponds to.
\item
\texttt{st4Meo.M11, .M12, .M21, .M22}: matrix elements of each of the four blocks.
\item
\texttt{st4Meo.ind1, .ind2}: row and column indices included in each of the four blocks.
\item
The full matrix can be reconstructed as follows:
\begin{align*}
\mathbf{M^{eo}} =\begin{pmatrix}
\texttt{M11(ind1,ind1)} & \texttt{M12(ind1,ind2)}\\[0.5cm]
\texttt{M21(ind2,ind1)} & \texttt{M22(ind2,ind2)}
\end{pmatrix}
\end{align*}
where each block is a \texttt{N+1-m x N+1-m} square matrix.
\end{itemize}

To obtain the full matrices in standard form, use the following call\\
\texttt{Qm = rvhGetFullMatrix(CstPQa\{m+1\},\textquotesingle st4MQ\textquotesingle);\\}

Note that functions that exploit this symmetry and the $oe-eo$ matrices are prefixed with \texttt{rvh}.

\subsection[Storage of $(n,m)$ arrays]{Storage of expansion coefficients and other $(n,m)$ arrays}

Several arrays depend on $(n,m)$, for example, the angular functions $\pi_{nm}$ and $\tau_{nm}$, the field expansion coefficients $a_{nm}$, etc. To store those arrays, we use the ``$p$-index'', which combines the possible values of $(n,m)$ in a linear array using the following convention $p = n(n+1) + m$. For a given maximum $N$, i.e. $1\le n \le N$ and $|m|\le n$, the length of the $p$-vectors is $P = N(N + 2)$.



\section*{Acknowledgments}
We acknowledge the support of the Royal Society of New Zealand (RSNZ) through a Marsden Grant (VUW1107) and Rutherford Discovery Fellowship (VUW1002).

\section{References}
\bibliographystyle{elsarticle-num}
\bibliography{Tmatrix}

\section{Appendix: Convergence tests}
\label{sec:tests}

\input{summaryXmaxProl.tex}

Table~\ref{tab:xmaxpro} summarises the results of semi-automated convergence tests for prolate spheroids with seven different values of refractive index. Each entry corresponds to the estimated largest $x_\text{max}=k_1\text{max}(a,c)$ (maximum size parameter) that the code can accurately model, for a given aspect ratio $h$, varied from 1.1 to 100. Almost identical limits are obtained for prolate and oblate spheroids, but one should note that those would be different
if expressed in terms of equivalent-volume-sphere radius, $r_V$.

Tables~\ref{tab:oblate} and \ref{tab:prolate} summarise the convergence parameters for oblate and prolate spheroids, respectively, with $s=1.311+0i$. For each combination of maximum size parameter $x_\text{max}$ and aspect ratio $h$, we list the pair of integers $N$ and $N_\theta$ that provided the optimum accuracy, judged from the convergence of the orientation-averaged extinction cross-section with increasing $N, N_\theta$. In some cases, the automated search failed to reach the optimum region of convergence, and the results presented in those tables should be considered with caution. We recommend users run their own convergence tests, using those tabulated values as general guidelines.

\sisetup{detect-all, retain-unity-mantissa = false,
tight-spacing=false,
scientific-notation=true, exponent-product = \cdot,
round-mode = figures,
round-precision = 0}
\setlength\tabcolsep{2pt}
\begin{landscape}
\input{summarylandObl.tex}
\end{landscape}
\begin{landscape}
\input{summarylandPro.tex}
\end{landscape}

\end{document}

%% file: summaryXmaxObl.tex
\sisetup{detect-all, tight-spacing=false, exponent-product = \cdot,
round-mode = off}
\begin{table}[!hbpt]
  \centering
  \scalebox{1}{
  \begin{tabular}{crrrrrr}
    \toprule[0.15em]
  &    \multicolumn{6}{c}{$h\rightarrow$}   \\
    \cmidrule(lr{0em}){2-7}
$s$ &  1.1 & 2 & 4   & 10 & 20 & 100 \\
    \midrule
  1.311 + 0.00i &   80 &  50 &  45 &  35 &  30 &  27 \\
     \midrule
  1.500 + 0.00i &   50 &  35 &  30 &  25 &  22 &  25 \\
     \midrule
  1.500 + 0.02i &   60 &  40 &  30 &  25 &  22 &  25 \\
     \midrule
  1.500 + 2.00i &   80 &  20 &  12 &  11 &  11 &   9 \\
     \midrule
  2.500 + 0.00i &   22 &  16 &  12 &  11 &  11 &  11 \\
     \midrule
  4.000 + 0.10i &   16 &  11 &   8 &   6 &   6 &   6 \\
     \midrule
  0.100 + 4.00i &   60 &  10 &   7 &   5 &   5 &   5 \\
     \bottomrule[0.15em]
  \end{tabular}
  }
\caption{Convergence study for oblate spheroids. We here consider a number of aspect ratios $h$ ranging from 1.1 to 100, and 7 representative values of $s$. For each, we calculate the orientation-averaged extinction cross-section for increasing sizes, characterized by the maximum size parameter $x_\text{max}=k_1\text{max}(a,c)$. The values in the table correspond to the largest $x_\text{max}$ for which convergence was obtained.
Those values are only indicators of the range of validity of the code; they were obtained via an automated search, which may be slightly inaccurate in some cases.}
\labelxmaxobl
\end{table}

%% file: summaryXmaxProl.tex
\sisetup{detect-all, tight-spacing=false, exponent-product = \cdot,
round-mode = off}
\begin{table}[!hbpt]
\centering
\scalebox{1}{
\begin{tabular}{crrrrrr}
  \toprule[0.15em]
&    \multicolumn{6}{c}{$h\rightarrow$}   \\
  \cmidrule(lr{0em}){2-7}
$s$ &  1.1 & 2 & 4 & 10 & 20 & 100 \\
  \midrule
1.311 + 0.00i &  80 &  50 &  45 &  35 &  35 &  35 \\
   \midrule
1.500 + 0.00i &  50 &  35 &  30 &  27 &  27 &  25 \\
   \midrule
1.500 + 0.02i &  60 &  40 &  30 &  27 &  27 &  27 \\
   \midrule
1.500 + 2.00i &  80 &  20 &  14 &  12 &  12 &  10 \\
   \midrule
2.500 + 0.00i &  20 &  18 &  12 &  12 &  12 &  12 \\
   \midrule
4.000 + 0.10i &  16 &  11 &   8 &   7 &   7 &   6 \\
   \midrule
0.100 + 4.00i & 60 &  10 &   7 &   6 &   6 &   5 \\
   \bottomrule[0.15em]
\end{tabular}
}
\caption{Convergence study for prolate spheroids. We here consider a number of aspect ratios $h$ ranging from 1.1 to 100, and 7 representative values of $s$. For each, we calculate the orientation-averaged extinction cross-section for increasing sizes, characterized by the maximum size parameter $x_\text{max}=k_1\text{max}(a,c)$. The values in the table correspond to the largest $x_\text{max}$ for which convergence was obtained.
Those values are only indicators of the range of validity of the code; they were obtained via an automated search, which may be slightly inaccurate in some cases.}
\labelxmaxpro
\end{table}

%% file: summarylandObl.tex
\begin{table}[!htpb]
\centering
\begin{tabular}{lllllllllllllllllllllll}
  \toprule[0.15em] 
 $h$: & $x_\text{max}$: & 0.01 & 0.1 & 1 & 2 & 3 & 4 & 5 & 6 & 7 & 8 & 10 & 12 & 15 & 20 & 25 & 30 & 40 & 50 & 60 & 70 & 80 \\ 
  \midrule
1.1 & \begin{tabular}{@{}l@{}}$N$ \\ $N_\theta$ \\ error \end{tabular} & \begin{tabular}{@{}l@{}}5 \\ 6 \\ -- \end{tabular} & \begin{tabular}{@{}l@{}}5 \\ 6 \\ -- \end{tabular} & \begin{tabular}{@{}l@{}}7 \\ 6 \\ -- \end{tabular} & \begin{tabular}{@{}l@{}}9 \\ 7 \\ -- \end{tabular} & \begin{tabular}{@{}l@{}}11 \\ 7 \\ -- \end{tabular} & \begin{tabular}{@{}l@{}}11 \\ 8 \\ -- \end{tabular} & \begin{tabular}{@{}l@{}}13 \\ 8 \\ -- \end{tabular} & \begin{tabular}{@{}l@{}}15 \\ 9 \\ -- \end{tabular} & \begin{tabular}{@{}l@{}}17 \\ 9 \\ -- \end{tabular} & \begin{tabular}{@{}l@{}}17 \\ 10 \\ -- \end{tabular} & \begin{tabular}{@{}l@{}}21 \\ 15 \\ -- \end{tabular} & \begin{tabular}{@{}l@{}}23 \\ 15 \\ -- \end{tabular} & \begin{tabular}{@{}l@{}}27 \\ 15 \\ -- \end{tabular} & \begin{tabular}{@{}l@{}}35 \\ 20 \\ -- \end{tabular} & \begin{tabular}{@{}l@{}}41 \\ 25 \\ -- \end{tabular} & \begin{tabular}{@{}l@{}}47 \\ 35 \\ -- \end{tabular} & \begin{tabular}{@{}l@{}}63 \\ 40 \\ \num{e-12} \end{tabular} & \begin{tabular}{@{}l@{}}77 \\ 40 \\ \num{e-11} \end{tabular} & \begin{tabular}{@{}l@{}}95 \\ 80 \\ \num{e-09} \end{tabular} & \begin{tabular}{@{}l@{}}127 \\ 80 \\ \num{e-07} \end{tabular} & \begin{tabular}{@{}l@{}}111 \\ 1200 \\ \num{e-04} \end{tabular} \\ 
   \midrule
1.3 & \begin{tabular}{@{}l@{}}$N$ \\ $N_\theta$ \\ error \end{tabular} & \begin{tabular}{@{}l@{}}5 \\ 9 \\ -- \end{tabular} & \begin{tabular}{@{}l@{}}5 \\ 9 \\ -- \end{tabular} & \begin{tabular}{@{}l@{}}9 \\ 9 \\ -- \end{tabular} & \begin{tabular}{@{}l@{}}11 \\ 9 \\ -- \end{tabular} & \begin{tabular}{@{}l@{}}11 \\ 9 \\ -- \end{tabular} & \begin{tabular}{@{}l@{}}13 \\ 10 \\ -- \end{tabular} & \begin{tabular}{@{}l@{}}15 \\ 15 \\ -- \end{tabular} & \begin{tabular}{@{}l@{}}15 \\ 15 \\ -- \end{tabular} & \begin{tabular}{@{}l@{}}19 \\ 15 \\ -- \end{tabular} & \begin{tabular}{@{}l@{}}19 \\ 15 \\ -- \end{tabular} & \begin{tabular}{@{}l@{}}23 \\ 15 \\ -- \end{tabular} & \begin{tabular}{@{}l@{}}27 \\ 20 \\ -- \end{tabular} & \begin{tabular}{@{}l@{}}31 \\ 20 \\ -- \end{tabular} & \begin{tabular}{@{}l@{}}37 \\ 25 \\ -- \end{tabular} & \begin{tabular}{@{}l@{}}45 \\ 45 \\ -- \end{tabular} & \begin{tabular}{@{}l@{}}53 \\ 30 \\ \num{e-12} \end{tabular} & \begin{tabular}{@{}l@{}}69 \\ 50 \\ \num{e-09} \end{tabular} & \begin{tabular}{@{}l@{}}93 \\ 45 \\ \num{e-07} \end{tabular} & \begin{tabular}{@{}l@{}}115 \\ 500 \\ \num{e-03} \end{tabular} &  &  \\ 
   \midrule
2 & \begin{tabular}{@{}l@{}}$N$ \\ $N_\theta$ \\ error \end{tabular} & \begin{tabular}{@{}l@{}}5 \\ 20 \\ -- \end{tabular} & \begin{tabular}{@{}l@{}}7 \\ 20 \\ -- \end{tabular} & \begin{tabular}{@{}l@{}}11 \\ 20 \\ -- \end{tabular} & \begin{tabular}{@{}l@{}}13 \\ 20 \\ -- \end{tabular} & \begin{tabular}{@{}l@{}}15 \\ 20 \\ -- \end{tabular} & \begin{tabular}{@{}l@{}}15 \\ 20 \\ -- \end{tabular} & \begin{tabular}{@{}l@{}}19 \\ 20 \\ -- \end{tabular} & \begin{tabular}{@{}l@{}}19 \\ 20 \\ -- \end{tabular} & \begin{tabular}{@{}l@{}}21 \\ 25 \\ -- \end{tabular} & \begin{tabular}{@{}l@{}}23 \\ 25 \\ -- \end{tabular} & \begin{tabular}{@{}l@{}}25 \\ 25 \\ -- \end{tabular} & \begin{tabular}{@{}l@{}}27 \\ 25 \\ -- \end{tabular} & \begin{tabular}{@{}l@{}}33 \\ 30 \\ -- \end{tabular} & \begin{tabular}{@{}l@{}}41 \\ 35 \\ -- \end{tabular} & \begin{tabular}{@{}l@{}}49 \\ 45 \\ \num{e-12} \end{tabular} & \begin{tabular}{@{}l@{}}61 \\ 35 \\ \num{e-11} \end{tabular} & \begin{tabular}{@{}l@{}}75 \\ 50 \\ \num{e-07} \end{tabular} & \begin{tabular}{@{}l@{}}103 \\ 45 \\ \num{e-04} \end{tabular} &  &  &  \\ 
   \midrule
4 & \begin{tabular}{@{}l@{}}$N$ \\ $N_\theta$ \\ error \end{tabular} & \begin{tabular}{@{}l@{}}5 \\ 40 \\ -- \end{tabular} & \begin{tabular}{@{}l@{}}7 \\ 40 \\ -- \end{tabular} & \begin{tabular}{@{}l@{}}11 \\ 40 \\ -- \end{tabular} & \begin{tabular}{@{}l@{}}15 \\ 40 \\ -- \end{tabular} & \begin{tabular}{@{}l@{}}17 \\ 40 \\ -- \end{tabular} & \begin{tabular}{@{}l@{}}19 \\ 7 \\ \num{e-03} \end{tabular} & \begin{tabular}{@{}l@{}}21 \\ 40 \\ -- \end{tabular} & \begin{tabular}{@{}l@{}}23 \\ 45 \\ -- \end{tabular} & \begin{tabular}{@{}l@{}}23 \\ 45 \\ -- \end{tabular} & \begin{tabular}{@{}l@{}}25 \\ 45 \\ -- \end{tabular} & \begin{tabular}{@{}l@{}}29 \\ 50 \\ -- \end{tabular} & \begin{tabular}{@{}l@{}}31 \\ 50 \\ -- \end{tabular} & \begin{tabular}{@{}l@{}}35 \\ 60 \\ -- \end{tabular} & \begin{tabular}{@{}l@{}}45 \\ 70 \\ \num{e-13} \end{tabular} & \begin{tabular}{@{}l@{}}51 \\ 70 \\ \num{e-11} \end{tabular} & \begin{tabular}{@{}l@{}}57 \\ 90 \\ \num{e-09} \end{tabular} & \begin{tabular}{@{}l@{}}59 \\ 90 \\ \num{e-04} \end{tabular} &  &  &  &  \\ 
   \midrule
7 & \begin{tabular}{@{}l@{}}$N$ \\ $N_\theta$ \\ error \end{tabular} & \begin{tabular}{@{}l@{}}5 \\ 70 \\ -- \end{tabular} & \begin{tabular}{@{}l@{}}7 \\ 70 \\ -- \end{tabular} & \begin{tabular}{@{}l@{}}11 \\ 70 \\ -- \end{tabular} & \begin{tabular}{@{}l@{}}15 \\ 70 \\ -- \end{tabular} & \begin{tabular}{@{}l@{}}19 \\ 70 \\ -- \end{tabular} & \begin{tabular}{@{}l@{}}19 \\ 70 \\ -- \end{tabular} & \begin{tabular}{@{}l@{}}19 \\ 80 \\ -- \end{tabular} & \begin{tabular}{@{}l@{}}23 \\ 80 \\ -- \end{tabular} & \begin{tabular}{@{}l@{}}25 \\ 80 \\ -- \end{tabular} & \begin{tabular}{@{}l@{}}27 \\ 80 \\ -- \end{tabular} & \begin{tabular}{@{}l@{}}31 \\ 90 \\ -- \end{tabular} & \begin{tabular}{@{}l@{}}35 \\ 90 \\ -- \end{tabular} & \begin{tabular}{@{}l@{}}37 \\ 100 \\ -- \end{tabular} & \begin{tabular}{@{}l@{}}41 \\ 200 \\ \num{e-11} \end{tabular} & \begin{tabular}{@{}l@{}}47 \\ 130 \\ \num{e-10} \end{tabular} & \begin{tabular}{@{}l@{}}47 \\ 130 \\ \num{e-06} \end{tabular} &  &  &  &  &  \\ 
   \midrule
10 & \begin{tabular}{@{}l@{}}$N$ \\ $N_\theta$ \\ error \end{tabular} & \begin{tabular}{@{}l@{}}5 \\ 100 \\ -- \end{tabular} & \begin{tabular}{@{}l@{}}7 \\ 100 \\ -- \end{tabular} & \begin{tabular}{@{}l@{}}13 \\ 90 \\ -- \end{tabular} & \begin{tabular}{@{}l@{}}15 \\ 100 \\ -- \end{tabular} & \begin{tabular}{@{}l@{}}19 \\ 100 \\ -- \end{tabular} & \begin{tabular}{@{}l@{}}19 \\ 100 \\ -- \end{tabular} & \begin{tabular}{@{}l@{}}21 \\ 110 \\ -- \end{tabular} & \begin{tabular}{@{}l@{}}25 \\ 110 \\ -- \end{tabular} & \begin{tabular}{@{}l@{}}27 \\ 110 \\ -- \end{tabular} & \begin{tabular}{@{}l@{}}27 \\ 120 \\ -- \end{tabular} & \begin{tabular}{@{}l@{}}31 \\ 120 \\ -- \end{tabular} & \begin{tabular}{@{}l@{}}33 \\ 130 \\ -- \end{tabular} & \begin{tabular}{@{}l@{}}41 \\ 140 \\ -- \end{tabular} & \begin{tabular}{@{}l@{}}45 \\ 180 \\ \num{e-11} \end{tabular} & \begin{tabular}{@{}l@{}}43 \\ 180 \\ \num{e-08} \end{tabular} & \begin{tabular}{@{}l@{}}47 \\ 200 \\ \num{e-06} \end{tabular} &  &  &  &  &  \\ 
   \midrule
20 & \begin{tabular}{@{}l@{}}$N$ \\ $N_\theta$ \\ error \end{tabular} & \begin{tabular}{@{}l@{}}5 \\ 200 \\ -- \end{tabular} & \begin{tabular}{@{}l@{}}7 \\ 200 \\ -- \end{tabular} & \begin{tabular}{@{}l@{}}13 \\ 200 \\ -- \end{tabular} & \begin{tabular}{@{}l@{}}15 \\ 200 \\ -- \end{tabular} & \begin{tabular}{@{}l@{}}19 \\ 200 \\ -- \end{tabular} & \begin{tabular}{@{}l@{}}21 \\ 200 \\ -- \end{tabular} & \begin{tabular}{@{}l@{}}23 \\ 200 \\ -- \end{tabular} & \begin{tabular}{@{}l@{}}23 \\ 220 \\ -- \end{tabular} & \begin{tabular}{@{}l@{}}27 \\ 220 \\ -- \end{tabular} & \begin{tabular}{@{}l@{}}29 \\ 240 \\ -- \end{tabular} & \begin{tabular}{@{}l@{}}33 \\ 260 \\ -- \end{tabular} & \begin{tabular}{@{}l@{}}37 \\ 260 \\ -- \end{tabular} & \begin{tabular}{@{}l@{}}43 \\ 280 \\ -- \end{tabular} & \begin{tabular}{@{}l@{}}43 \\ 300 \\ \num{e-12} \end{tabular} & \begin{tabular}{@{}l@{}}49 \\ 550 \\ \num{e-08} \end{tabular} & \begin{tabular}{@{}l@{}}41 \\ 400 \\ \num{e-04} \end{tabular} &  &  &  &  &  \\ 
   \midrule
50 & \begin{tabular}{@{}l@{}}$N$ \\ $N_\theta$ \\ error \end{tabular} & \begin{tabular}{@{}l@{}}5 \\ 500 \\ -- \end{tabular} & \begin{tabular}{@{}l@{}}7 \\ 500 \\ -- \end{tabular} & \begin{tabular}{@{}l@{}}13 \\ 500 \\ -- \end{tabular} & \begin{tabular}{@{}l@{}}17 \\ 500 \\ -- \end{tabular} & \begin{tabular}{@{}l@{}}19 \\ 500 \\ -- \end{tabular} & \begin{tabular}{@{}l@{}}21 \\ 500 \\ -- \end{tabular} & \begin{tabular}{@{}l@{}}23 \\ 550 \\ -- \end{tabular} & \begin{tabular}{@{}l@{}}25 \\ 550 \\ -- \end{tabular} & \begin{tabular}{@{}l@{}}27 \\ 550 \\ -- \end{tabular} & \begin{tabular}{@{}l@{}}29 \\ 600 \\ -- \end{tabular} & \begin{tabular}{@{}l@{}}33 \\ 650 \\ -- \end{tabular} & \begin{tabular}{@{}l@{}}35 \\ 800 \\ -- \end{tabular} & \begin{tabular}{@{}l@{}}39 \\ 700 \\ \num{e-13} \end{tabular} & \begin{tabular}{@{}l@{}}45 \\ 1100 \\ \num{e-12} \end{tabular} & \begin{tabular}{@{}l@{}}47 \\ 900 \\ \num{e-09} \end{tabular} & \begin{tabular}{@{}l@{}}45 \\ 1200 \\ \num{e-06} \end{tabular} &  &  &  &  &  \\ 
   \midrule
100 & \begin{tabular}{@{}l@{}}$N$ \\ $N_\theta$ \\ error \end{tabular} & \begin{tabular}{@{}l@{}}5 \\ 1000 \\ -- \end{tabular} & \begin{tabular}{@{}l@{}}7 \\ 1100 \\ -- \end{tabular} & \begin{tabular}{@{}l@{}}13 \\ 1000 \\ -- \end{tabular} & \begin{tabular}{@{}l@{}}17 \\ 1100 \\ -- \end{tabular} & \begin{tabular}{@{}l@{}}19 \\ 1000 \\ -- \end{tabular} & \begin{tabular}{@{}l@{}}21 \\ 1100 \\ -- \end{tabular} & \begin{tabular}{@{}l@{}}23 \\ 1100 \\ -- \end{tabular} & \begin{tabular}{@{}l@{}}25 \\ 1100 \\ -- \end{tabular} & \begin{tabular}{@{}l@{}}27 \\ 1100 \\ -- \end{tabular} & \begin{tabular}{@{}l@{}}27 \\ 1100 \\ -- \end{tabular} & \begin{tabular}{@{}l@{}}33 \\ 1300 \\ -- \end{tabular} & \begin{tabular}{@{}l@{}}35 \\ 1500 \\ -- \end{tabular} & \begin{tabular}{@{}l@{}}47 \\ 1400 \\ -- \end{tabular} & \begin{tabular}{@{}l@{}}53 \\ 1600 \\ -- \end{tabular} & \begin{tabular}{@{}l@{}}47 \\ 2000 \\ \num{e-09} \end{tabular} &  &  &  &  &  &  \\ 
   \bottomrule[0.15em] 
\end{tabular}
\caption{Convergence study for an  oblate  spheroid with $s= 1.311+0i $.
  For each pair of size parameter $x_\text{max}$ (columns) and aspect ratio $h$ (rows), 
  we list the convergence parameters $N, N_\theta$, together with the relative error, only displayed if worse than $10^{-13}$.
} 
\label{tab:oblate}
\end{table}

%% file: summarylandPro.tex
\begin{table}[!htpb]
\centering
\begin{tabular}{lllllllllllllllllllllll}
  \toprule[0.15em] 
 $h$: & $x_\text{max}$: & 0.01 & 0.1 & 1 & 2 & 3 & 4 & 5 & 6 & 7 & 8 & 10 & 12 & 15 & 20 & 25 & 30 & 40 & 50 & 60 & 70 & 80 \\ 
  \midrule
1.1 & \begin{tabular}{@{}l@{}}$N$ \\ $N_\theta$ \\ error \end{tabular} & \begin{tabular}{@{}l@{}}5 \\ 6 \\ -- \end{tabular} & \begin{tabular}{@{}l@{}}5 \\ 6 \\ -- \end{tabular} & \begin{tabular}{@{}l@{}}7 \\ 6 \\ -- \end{tabular} & \begin{tabular}{@{}l@{}}9 \\ 7 \\ -- \end{tabular} & \begin{tabular}{@{}l@{}}11 \\ 7 \\ -- \end{tabular} & \begin{tabular}{@{}l@{}}11 \\ 8 \\ -- \end{tabular} & \begin{tabular}{@{}l@{}}13 \\ 8 \\ -- \end{tabular} & \begin{tabular}{@{}l@{}}15 \\ 9 \\ -- \end{tabular} & \begin{tabular}{@{}l@{}}17 \\ 10 \\ -- \end{tabular} & \begin{tabular}{@{}l@{}}17 \\ 10 \\ -- \end{tabular} & \begin{tabular}{@{}l@{}}21 \\ 15 \\ -- \end{tabular} & \begin{tabular}{@{}l@{}}23 \\ 15 \\ -- \end{tabular} & \begin{tabular}{@{}l@{}}27 \\ 15 \\ -- \end{tabular} & \begin{tabular}{@{}l@{}}35 \\ 20 \\ -- \end{tabular} & \begin{tabular}{@{}l@{}}41 \\ 25 \\ -- \end{tabular} & \begin{tabular}{@{}l@{}}47 \\ 25 \\ -- \end{tabular} & \begin{tabular}{@{}l@{}}67 \\ 40 \\ \num{e-12} \end{tabular} & \begin{tabular}{@{}l@{}}79 \\ 60 \\ \num{e-10} \end{tabular} & \begin{tabular}{@{}l@{}}97 \\ 45 \\ \num{e-09} \end{tabular} & \begin{tabular}{@{}l@{}}87 \\ 70 \\ \num{e-07} \end{tabular} & \begin{tabular}{@{}l@{}}109 \\ 120 \\ \num{e-04} \end{tabular} \\ 
   \midrule
1.3 & \begin{tabular}{@{}l@{}}$N$ \\ $N_\theta$ \\ error \end{tabular} & \begin{tabular}{@{}l@{}}5 \\ 9 \\ -- \end{tabular} & \begin{tabular}{@{}l@{}}5 \\ 9 \\ -- \end{tabular} & \begin{tabular}{@{}l@{}}9 \\ 9 \\ -- \end{tabular} & \begin{tabular}{@{}l@{}}11 \\ 9 \\ -- \end{tabular} & \begin{tabular}{@{}l@{}}11 \\ 9 \\ -- \end{tabular} & \begin{tabular}{@{}l@{}}13 \\ 10 \\ -- \end{tabular} & \begin{tabular}{@{}l@{}}15 \\ 15 \\ -- \end{tabular} & \begin{tabular}{@{}l@{}}15 \\ 15 \\ -- \end{tabular} & \begin{tabular}{@{}l@{}}19 \\ 15 \\ -- \end{tabular} & \begin{tabular}{@{}l@{}}19 \\ 15 \\ -- \end{tabular} & \begin{tabular}{@{}l@{}}23 \\ 15 \\ -- \end{tabular} & \begin{tabular}{@{}l@{}}27 \\ 20 \\ -- \end{tabular} & \begin{tabular}{@{}l@{}}31 \\ 20 \\ -- \end{tabular} & \begin{tabular}{@{}l@{}}37 \\ 25 \\ -- \end{tabular} & \begin{tabular}{@{}l@{}}45 \\ 25 \\ -- \end{tabular} & \begin{tabular}{@{}l@{}}53 \\ 30 \\ \num{e-12} \end{tabular} & \begin{tabular}{@{}l@{}}73 \\ 40 \\ \num{e-10} \end{tabular} & \begin{tabular}{@{}l@{}}95 \\ 45 \\ \num{e-07} \end{tabular} & \begin{tabular}{@{}l@{}}117 \\ 180 \\ \num{e-04} \end{tabular} &  &  \\ 
   \midrule
2 & \begin{tabular}{@{}l@{}}$N$ \\ $N_\theta$ \\ error \end{tabular} & \begin{tabular}{@{}l@{}}5 \\ 20 \\ -- \end{tabular} & \begin{tabular}{@{}l@{}}7 \\ 20 \\ -- \end{tabular} & \begin{tabular}{@{}l@{}}11 \\ 20 \\ -- \end{tabular} & \begin{tabular}{@{}l@{}}13 \\ 20 \\ -- \end{tabular} & \begin{tabular}{@{}l@{}}15 \\ 20 \\ -- \end{tabular} & \begin{tabular}{@{}l@{}}15 \\ 20 \\ -- \end{tabular} & \begin{tabular}{@{}l@{}}19 \\ 20 \\ -- \end{tabular} & \begin{tabular}{@{}l@{}}19 \\ 20 \\ -- \end{tabular} & \begin{tabular}{@{}l@{}}21 \\ 20 \\ -- \end{tabular} & \begin{tabular}{@{}l@{}}23 \\ 25 \\ -- \end{tabular} & \begin{tabular}{@{}l@{}}25 \\ 25 \\ -- \end{tabular} & \begin{tabular}{@{}l@{}}27 \\ 25 \\ -- \end{tabular} & \begin{tabular}{@{}l@{}}33 \\ 30 \\ -- \end{tabular} & \begin{tabular}{@{}l@{}}41 \\ 30 \\ -- \end{tabular} & \begin{tabular}{@{}l@{}}49 \\ 45 \\ \num{e-12} \end{tabular} & \begin{tabular}{@{}l@{}}61 \\ 45 \\ \num{e-10} \end{tabular} & \begin{tabular}{@{}l@{}}89 \\ 40 \\ \num{e-07} \end{tabular} & \begin{tabular}{@{}l@{}}113 \\ 50 \\ \num{e-04} \end{tabular} &  &  &  \\ 
   \midrule
4 & \begin{tabular}{@{}l@{}}$N$ \\ $N_\theta$ \\ error \end{tabular} & \begin{tabular}{@{}l@{}}5 \\ 40 \\ -- \end{tabular} & \begin{tabular}{@{}l@{}}7 \\ 40 \\ -- \end{tabular} & \begin{tabular}{@{}l@{}}11 \\ 40 \\ -- \end{tabular} & \begin{tabular}{@{}l@{}}15 \\ 40 \\ -- \end{tabular} & \begin{tabular}{@{}l@{}}17 \\ 40 \\ -- \end{tabular} & \begin{tabular}{@{}l@{}}19 \\ 40 \\ -- \end{tabular} & \begin{tabular}{@{}l@{}}19 \\ 40 \\ -- \end{tabular} & \begin{tabular}{@{}l@{}}23 \\ 40 \\ -- \end{tabular} & \begin{tabular}{@{}l@{}}23 \\ 45 \\ -- \end{tabular} & \begin{tabular}{@{}l@{}}25 \\ 45 \\ -- \end{tabular} & \begin{tabular}{@{}l@{}}27 \\ 45 \\ -- \end{tabular} & \begin{tabular}{@{}l@{}}31 \\ 50 \\ -- \end{tabular} & \begin{tabular}{@{}l@{}}35 \\ 60 \\ -- \end{tabular} & \begin{tabular}{@{}l@{}}43 \\ 90 \\ -- \end{tabular} & \begin{tabular}{@{}l@{}}49 \\ 70 \\ \num{e-12} \end{tabular} & \begin{tabular}{@{}l@{}}49 \\ 100 \\ \num{e-09} \end{tabular} & \begin{tabular}{@{}l@{}}59 \\ 110 \\ \num{e-06} \end{tabular} &  &  &  &  \\ 
   \midrule
7 & \begin{tabular}{@{}l@{}}$N$ \\ $N_\theta$ \\ error \end{tabular} & \begin{tabular}{@{}l@{}}5 \\ 80 \\ -- \end{tabular} & \begin{tabular}{@{}l@{}}7 \\ 80 \\ -- \end{tabular} & \begin{tabular}{@{}l@{}}11 \\ 80 \\ -- \end{tabular} & \begin{tabular}{@{}l@{}}15 \\ 70 \\ -- \end{tabular} & \begin{tabular}{@{}l@{}}17 \\ 70 \\ -- \end{tabular} & \begin{tabular}{@{}l@{}}19 \\ 70 \\ -- \end{tabular} & \begin{tabular}{@{}l@{}}21 \\ 70 \\ -- \end{tabular} & \begin{tabular}{@{}l@{}}23 \\ 80 \\ -- \end{tabular} & \begin{tabular}{@{}l@{}}25 \\ 80 \\ -- \end{tabular} & \begin{tabular}{@{}l@{}}27 \\ 80 \\ -- \end{tabular} & \begin{tabular}{@{}l@{}}31 \\ 80 \\ -- \end{tabular} & \begin{tabular}{@{}l@{}}33 \\ 90 \\ -- \end{tabular} & \begin{tabular}{@{}l@{}}37 \\ 90 \\ -- \end{tabular} & \begin{tabular}{@{}l@{}}43 \\ 120 \\ \num{e-13} \end{tabular} & \begin{tabular}{@{}l@{}}43 \\ 120 \\ \num{e-10} \end{tabular} & \begin{tabular}{@{}l@{}}47 \\ 130 \\ \num{e-08} \end{tabular} &  &  &  &  &  \\ 
   \midrule
10 & \begin{tabular}{@{}l@{}}$N$ \\ $N_\theta$ \\ error \end{tabular} & \begin{tabular}{@{}l@{}}5 \\ 120 \\ -- \end{tabular} & \begin{tabular}{@{}l@{}}7 \\ 120 \\ -- \end{tabular} & \begin{tabular}{@{}l@{}}11 \\ 120 \\ -- \end{tabular} & \begin{tabular}{@{}l@{}}15 \\ 120 \\ -- \end{tabular} & \begin{tabular}{@{}l@{}}17 \\ 120 \\ -- \end{tabular} & \begin{tabular}{@{}l@{}}21 \\ 100 \\ -- \end{tabular} & \begin{tabular}{@{}l@{}}21 \\ 120 \\ -- \end{tabular} & \begin{tabular}{@{}l@{}}23 \\ 120 \\ -- \end{tabular} & \begin{tabular}{@{}l@{}}27 \\ 120 \\ -- \end{tabular} & \begin{tabular}{@{}l@{}}27 \\ 120 \\ -- \end{tabular} & \begin{tabular}{@{}l@{}}31 \\ 120 \\ -- \end{tabular} & \begin{tabular}{@{}l@{}}33 \\ 120 \\ -- \end{tabular} & \begin{tabular}{@{}l@{}}37 \\ 140 \\ -- \end{tabular} & \begin{tabular}{@{}l@{}}45 \\ 160 \\ -- \end{tabular} & \begin{tabular}{@{}l@{}}45 \\ 160 \\ \num{e-10} \end{tabular} & \begin{tabular}{@{}l@{}}45 \\ 200 \\ \num{e-08} \end{tabular} &  &  &  &  &  \\ 
   \midrule
20 & \begin{tabular}{@{}l@{}}$N$ \\ $N_\theta$ \\ error \end{tabular} & \begin{tabular}{@{}l@{}}5 \\ 220 \\ \num{e-13} \end{tabular} & \begin{tabular}{@{}l@{}}7 \\ 220 \\ \num{e-13} \end{tabular} & \begin{tabular}{@{}l@{}}11 \\ 220 \\ \num{e-13} \end{tabular} & \begin{tabular}{@{}l@{}}15 \\ 260 \\ -- \end{tabular} & \begin{tabular}{@{}l@{}}19 \\ 220 \\ -- \end{tabular} & \begin{tabular}{@{}l@{}}21 \\ 220 \\ -- \end{tabular} & \begin{tabular}{@{}l@{}}21 \\ 220 \\ -- \end{tabular} & \begin{tabular}{@{}l@{}}25 \\ 220 \\ -- \end{tabular} & \begin{tabular}{@{}l@{}}27 \\ 220 \\ -- \end{tabular} & \begin{tabular}{@{}l@{}}27 \\ 260 \\ -- \end{tabular} & \begin{tabular}{@{}l@{}}31 \\ 260 \\ -- \end{tabular} & \begin{tabular}{@{}l@{}}29 \\ 260 \\ \num{e-12} \end{tabular} & \begin{tabular}{@{}l@{}}39 \\ 280 \\ -- \end{tabular} & \begin{tabular}{@{}l@{}}47 \\ 300 \\ \num{e-13} \end{tabular} & \begin{tabular}{@{}l@{}}49 \\ 350 \\ \num{e-10} \end{tabular} & \begin{tabular}{@{}l@{}}45 \\ 350 \\ \num{e-07} \end{tabular} &  &  &  &  &  \\ 
   \midrule
50 & \begin{tabular}{@{}l@{}}$N$ \\ $N_\theta$ \\ error \end{tabular} & \begin{tabular}{@{}l@{}}5 \\ 650 \\ \num{e-12} \end{tabular} & \begin{tabular}{@{}l@{}}7 \\ 650 \\ \num{e-12} \end{tabular} & \begin{tabular}{@{}l@{}}11 \\ 650 \\ \num{e-12} \end{tabular} & \begin{tabular}{@{}l@{}}15 \\ 400 \\ \num{e-12} \end{tabular} & \begin{tabular}{@{}l@{}}19 \\ 500 \\ -- \end{tabular} & \begin{tabular}{@{}l@{}}21 \\ 500 \\ -- \end{tabular} & \begin{tabular}{@{}l@{}}21 \\ 500 \\ -- \end{tabular} & \begin{tabular}{@{}l@{}}25 \\ 500 \\ -- \end{tabular} & \begin{tabular}{@{}l@{}}27 \\ 650 \\ -- \end{tabular} & \begin{tabular}{@{}l@{}}29 \\ 700 \\ -- \end{tabular} & \begin{tabular}{@{}l@{}}31 \\ 600 \\ -- \end{tabular} & \begin{tabular}{@{}l@{}}33 \\ 700 \\ -- \end{tabular} & \begin{tabular}{@{}l@{}}39 \\ 700 \\ -- \end{tabular} & \begin{tabular}{@{}l@{}}45 \\ 900 \\ \num{e-12} \end{tabular} & \begin{tabular}{@{}l@{}}53 \\ 1100 \\ \num{e-11} \end{tabular} & \begin{tabular}{@{}l@{}}45 \\ 1200 \\ \num{e-07} \end{tabular} &  &  &  &  &  \\ 
   \midrule
100 & \begin{tabular}{@{}l@{}}$N$ \\ $N_\theta$ \\ error \end{tabular} & \begin{tabular}{@{}l@{}}5 \\ 800 \\ \num{e-12} \end{tabular} & \begin{tabular}{@{}l@{}}7 \\ 800 \\ \num{e-12} \end{tabular} & \begin{tabular}{@{}l@{}}11 \\ 800 \\ \num{e-12} \end{tabular} & \begin{tabular}{@{}l@{}}15 \\ 800 \\ \num{e-12} \end{tabular} & \begin{tabular}{@{}l@{}}19 \\ 800 \\ \num{e-12} \end{tabular} & \begin{tabular}{@{}l@{}}21 \\ 1000 \\ \num{e-12} \end{tabular} & \begin{tabular}{@{}l@{}}21 \\ 1000 \\ \num{e-13} \end{tabular} & \begin{tabular}{@{}l@{}}25 \\ 1000 \\ \num{e-13} \end{tabular} & \begin{tabular}{@{}l@{}}27 \\ 1000 \\ \num{e-13} \end{tabular} & \begin{tabular}{@{}l@{}}29 \\ 1400 \\ -- \end{tabular} & \begin{tabular}{@{}l@{}}31 \\ 1400 \\ -- \end{tabular} & \begin{tabular}{@{}l@{}}33 \\ 1200 \\ -- \end{tabular} & \begin{tabular}{@{}l@{}}39 \\ 1600 \\ -- \end{tabular} & \begin{tabular}{@{}l@{}}45 \\ 1800 \\ \num{e-12} \end{tabular} & \begin{tabular}{@{}l@{}}47 \\ 2000 \\ \num{e-11} \end{tabular} & \begin{tabular}{@{}l@{}}47 \\ 2000 \\ \num{e-08} \end{tabular} &  &  &  &  &  \\ 
   \bottomrule[0.15em] 
\end{tabular}
\caption{Convergence study for a  prolate  spheroid with $s= 1.311+0i $.
  For each pair of size parameter $x_\text{max}$ (columns) and aspect ratio $h$ (rows), 
  we list the convergence parameters $N, N_\theta$, together with the relative error, only displayed if worse than $10^{-13}$.
} 
\label{tab:prolate}
\end{table}